\documentclass[manuscript]{aastex}

\begin{document}

\newcommand{\mz}{z_{\rm{850}}}

\title{The Star Formation History of Redshift $z\sim2$ Galaxies: The Role of The Infrared Prior.}

   \author{Lulu Fan\altaffilmark{1,2,3}, Andrea Lapi\altaffilmark{4,3},
Alessandro Bressan\altaffilmark{3}, Mario Nonino\altaffilmark{5}, Gianfranco De Zotti\altaffilmark{3}, and
Luigi Danese\altaffilmark{3}}
\altaffiltext{1} {Center for Astrophysics, University of Science and Technology of China, 230026 Hefei, China; {\it llfan@ustc.edu.cn}}
\altaffiltext{2}{Key Laboratory for Research in Galaxies and Cosmology, University of Science and Technology of China, Chinese Academy of Sciences, Hefei, Anhui, 230026, China}
\altaffiltext{3}{Astrophysics Sector, SISSA, Via Bonomea 265, 34136 Trieste, Italy}
\altaffiltext{4}{Dip. Fisica, Univ. `Tor Vergata', Via Ricerca Scientifica 1, 00133 Roma, Italy}
\altaffiltext{5}{INAF-Osservatorio Astronomico di Trieste, Via G.B. Tiepolo 11, 40131 Trieste, Italy}

\begin{abstract}
We build a sample of 298 spectroscopically-confirmed galaxies at redshift
$z\sim2$, selected in the $\mz$-band from the GOODS-MUSIC catalog.
By exploiting the rest frame 8$\mu m$ luminosity as a proxy of the star formation
rate (SFR) we check the accuracy of the standard SED-fitting technique, finding
it is not accurate enough to provide reliable estimates of the galaxy physical parameters.
We then develop a new SED-fitting method that includes the IR luminosity as a prior
and a generalized Calzetti law with a variable $R_V$ . Then we exploit such a new
method to re-analyze our galaxy sample, and to robustly determine SFRs, stellar masses
and ages. We find that there is a general trend of increasing attenuation with
the SFR. Moreover, we find that the SFRs range between a few
to $10^3M_\odot yr^{-1}$, the masses from $10^9$ to $4\times 10^{11}M_\odot$, while the ages from a few
tens of Myr to more than 1 Gyr. We discuss how individual age measurements of
highly attenuated objects indicate that dust must form within a few tens of Myr
and be copious already at  $\leq$100 Myr. In addition, we find that low luminous
galaxies harbor, on average, significantly older stellar populations and are
also less massive than brighter ones; we discuss how these findings and the well
known `downsizing' scenario are consistent in a framework where less massive
galaxies form first, but their star formation lasts longer. Finally, we find that the
near-IR attenuation is not scarce for luminous objects, contrary to what is customarily
assumed; we discuss how this affects the interpretation of the observed
$M_\star/L$ ratios.
\end{abstract}


%
\section{Introduction}           
\label{sect:intro}

Dust plays a crucial role in the formation and evolution of
galaxies. It absorbs stellar light and re-emits in far infrared
(FIR). Even a small amount of dust can lead to a significant
underestimation of SFR. Poggianti \& Wu (2000), Poggianti, Bressan, \& Franceschini (2001) and Rigopoulou et
al. (2000) reported independent evidence on both local and high-redshift
luminous starbursts in which $\sim$ 70\%-80\% of the bolometric flux
from young stars is completely obscured by dust and remains hidden
in the UV/optical surveys (even after correction for dust).

Dust constitutes a fundamental physical component in  active star-forming
galaxies. In fact, the dust formation is closely related to the
star formation activity, as it occurs in a range of environments
, from explosive ejecta of supernovae to the outflows of
evolved low-mass stars (Dwek 1998). Moreover, the presence of dust can
enhance the SFR (Hirashita \& Ferrara 2002; Morgan \& Edmunds
2003), as the surface of dust grains constitutes a site for efficient
formation of $H_2$ molecules (e.g. Cazaux \& Tielens 2002, 2004), which
act as an effective coolant in metal-poor interstellar media (ISM). In fact,
such a correlation has been found in both low and high redshift galaxies
(Adelberger \& Steidel 2000; Vijh et al. 2003; Ouchi et al. 2004; Shapley et al. 2001,2005).

To represent dust attenuation, most of the authors use UV slope $\beta_{UV}$,
and adopt the Meurer relation as derived from a sample of local starburst
galaxies (Meurer et al. 1999); other authors use instead the color excess E(B-V) as inferred from
the spectral energy distribution (SED) of galaxies, and adopt a Calzetti extinction law similar to the Meurer relation (Calzetti et al. 2000).
However, not only the validity of the local Meurer relation at high redshift
is currently under debate, but even in local
normal star-forming galaxies (i.e. Bell 2002; Seibert et al. 2005; Boissier
et al. 2007; Buat et al. 2005; Burgarella et al. 2005; Gil de Paz et
al. 2007; Cortese et al. 2006) and individual HII regions (Calzetti et al. 2005),
there are clear deviations from it. The most useful
indicator of dust obscuration should be infrared to ultraviolet luminosity
ratio ($IRX$),  which is relatively independent on star formation history
(Buat et al. 2005) and on the dust configuration and distribution
(Witt \& Gordon 2000).

Unfortunately, it's difficult to trace
the infrared luminosity of typical star forming galaxies at
high redshift directly by their far-infrared or submillimeter
emission, due to current sensitivity limits of bolometers and
submillimeter interferometers. We must look for other portions
of SED to examine dust emissions. A possible way is to use the
mid-infrared (mid-IR) dust emission of galaxies as a tracer of total
infrared luminosity. The correlation between mid-IR and infrared
luminosity has been revealed in both the local and $z\sim 1$ universe
with the observations of the \emph{Infrared Space Observatory(ISO)}
(Boselli et al. 1998; Adelberger \& Steidel 2000; Dale et al. 2000;
Helou et al. 2000; F\"{o}rster Schreiber et al. 2003) and the
Spitzer MIPS (Rieke et al. 2009). This relation has been also examined
at $z\sim 2$ by X-ray and radio stacking analysis
(Reddy et al. 2006) and Spitzer MIPS 24 $\mu m$ observations (Reddy et al. 2010).
Taking the $IRX$ derived from rest-frame 8$\mu m$ luminosity $L_{8}$
as a reference, one can test what extent the UV slope $\beta_{UV}$ is
an effective probe of the dust attenuation.

Another fundamental physical measurement for high redshift galaxies
is constituted by the stellar mass. Stellar population synthesis modeling is popularly
used to estimate stellar mass at high redshift (e.g. Sawicki \& Yee 1998;
Papovich et al. 2001, 2004, 2006; Shapley et al. 2005;
F\"{o}rster Schreiber et al. 2004). In particular, a number of authors in
recent years have interpreted observed-frame UV to near-IR (even mid-IR)
photometry with stellar population synthesis models in order to infer stellar masses
and the total mass as a function of redshift (e.g. Papovich et al. 2001;
Shapley et al.2001, 2005; Daddi et al.2004; Dickinson et al. 2003; Fontana et al. 2003;
Labb\'{e} et al. 2005;P\'{e}rez-Gonz\'{a}lez et al. 2008).
At $z\sim 2$, $H_{\alpha}+[NII]$ line emissions will be moved into
$K_S$ band, one has to take nebular spectrum into account in order to obtain precise
evaluations of stellar masses. We stress that, although Papovich et al.(2001) has tested
the impact of different metallicities, initial mass functions (IMFs) and
star formation histories on stellar mass determinations, the effects of assumptions concerning the dust
extinction law have to be checked in detail.

In this paper, we investigate dust attenuations, star formation histories and stellar
masses in $z\sim2$ galaxies. To this purpose, we exploit a data set spanning the e.m. spectrum
from UV to mid-IR, supplemented by Spitzer MIPS 24$\mu m$ photometry and spectroscopic
redshift determinations. We aim at showing that, in order to robustly evaluate the
star formation history and stellar mass at high redshift, the rest-frame SEDs from UV to
near-IR must be complemented by the IR luminosity, or at least by a robust prior for it.

The plan of the paper is as follows: in \S\ 2 we describe our sample selection; in \S\ 3 we
discuss the relation between the rest-frame UV slope and dust attenuation; in \S\ 4 and in the
Appendix we give details on our stellar population synthesis model; in \S\ 5 we present our results;
in \S\ 6 we summarize our main conclusions. Throughout the paper we assume a concordance $\Lambda$CDM cosmology with $\Omega_m=0.27$,
$\Omega_{\Lambda}=0.73$, and \emph{h}=0.71.

\section{The Sample Selection}

The data are taken directly from the Version 2 release of the
GOODS-MUSIC(GOODS MUlticolour Southern Infrared Catalog) sample
(Santini et al. 2009). The main differences between the Version 1
(Grazian et al. 2006) and the Version 2 of GOODS-MUSIC sample have
basically three points: (1) The IRAC photometry are improved
using new IRAC PSFs and background subtraction for the 3.6$\mu m$, 4.5$\mu m$, 5.8$\mu m$
and 8.0 $\mu m$ bands; (2) New spectroscopic redshifts available after
the Version 1 release are added; (3) The 24$\mu m$ photometry by the
MIPS instrument are added.

We start our work from the $\mz$-selected
catalog which has a typical $\mz$-band magnitude limit, $\mz$=26.0, for
most of the catalog and extends down to $\mz$=26.18 in limited
areas. The galaxies are selected from the $\mz$-selected catalog with
the flag $selgal$ = 1. Thanks to the releases of new spectra in the GOODS-South Field
(FORS2: Vanzella et al. 2006, 2008; VIMOS: Popesso et al. 2009, Balestra et al. 2010), we can
use a much larger spectroscopic sample than that presented in
Grazain et al. (2006). The quality of all spectroscopic redshifts is
marked with a $quality\_flag$ flag which is divided into four
classes: very good, good, uncertain and very uncertain, represented by
0, 1, 2 and 3, respectively. We discard those galaxies with uncertain
spectroscopic redshifts ($quality\_flag$ =\ 2 or \ 3).

The photometric measurements used to model the stellar populations
are comprised by 12 bands, namely U-band from the 2.2ESO ($U_{35}$ and
$U_{38}$) and VLT-VIMOS ($U_{VIMOS}$), the {\it F435W} (B-band),  {\it
F606W} (V-band), {\it F775W} (i-band), and {\it F850LP} ($\mz$-band) HST/ACS
data, the JH$K_S$ VLT data, and the 3.6$\mu m$, 4.5$\mu m$, 5.8$\mu m$ and 8.0 $\mu m$
bands from Spitzer IRAC instrument. We use $U_{VIMOS}$ instead of
$U_{35}$ and $U_{38}$ because the former is much deeper than the
other two. We select only the galaxies with at least one of JH$K_S$
bands, at least one of four IRAC bands and at least seven bands detected in total.

In this way we get a spectroscopic sample of 298 $\mz$-selected galaxies within the redshift range $1.4 \leq z \leq 3.0$.
We notice that most of our low redshift galaxies at $z\leq 1.7$ are
taken from FORS2 spectra which have red $i-\mz$ colors
($i-\mz=0.35\pm 0.21$ in our sample).
Most of our high redshift galaxies at $z\geq 1.7$ are
taken from VIMOS spectra which have blue $i-\mz$ colors
($i-\mz=0.09\pm 0.17$ in our sample). Finally, public 24$\mu m$ Spitzer
MIPS observations have also been exploited. A PSF-matching
technique, which is performed by the software \texttt{ConvPhot} (De Santis et
al. 2007), has been employed to properly detect and deblend objects in the MIPS images (Santini et al.2009).
As a result, 135 out of 298 galaxies are also detected in 24$\mu m$ band.
We emphasize that galaxies in $\mz$-selected sample at $z\sim2$ are expected to be star forming as
the sample is basically rest-frame near-UV selected. And a diagnostic BzK diagram confirms this
idea: over 95\% galaxies are sBzk.

The existence of a tight correlation between the rest-frame 8 $\mu m$ luminosity $\mu L_{\mu}$ (hereafter
$L_8$) and the $8-1000\mu m$ IR luminosity (hereafter $L_{IR}$), initially established
for local star forming galaxies, has been confirmed for large galaxy samples in a wide range
of luminosity and redshift, thanks to the advent of the Herschel satellite (see Chary \& Elbaz
2001; Caputi et al. 2007; Dale et al. 2007; Bavouzet et al. 2008; Rieke et al. 2009; Elbaz et
al. 2011; Nordon et al. 2010, 2012; Reddy et al. 2010, 2012a). Although the dependence on
redshift and luminosity is very weak, it has been suggested that the most relevant parameter
shaping the above relationship is the projected star formation density, as inferred from the
IR surface brightness (Elbaz et al. 2011; Nordon et al. 2012; Reddy et al. 2012a). While
extended galaxies exhibit a median ratio $IR8 = L_{IR}/L_8 \approx5$, the median $IR8$ of the compact
objects is larger by a factor around 2. These altogether exhibit
a median $IR8\approx8.5$, in agreement with the finding of Rigby et al. (2008) for a sample of
lensed galaxies at $z\sim2$. In the following, we will assume the median value $IR8=8.5$ as
a reference for our sample, in order to estimate the IR luminosity from 24$\mu m$ observations.
Following Elbaz et al. (2011) this value will possibly overestimate the inferred $L_{IR}$
of normal star-forming galaxies while it will underestimate that of compact starbursts.

Since the median redshift of our sample is $z\sim2$, galaxies are typically selected in the
rest-frame U band with broad-band spectra from 1200$\AA$ to the near-IR. Moreover, for the
objects detected at 8$\mu m$ we can use the $IR8$ relation to estimate the IR luminosity. We
notice that at those redshifts a direct measure of the IR luminosity is achievable (e.g.,
with Herschel) only for SFR$\geq 10^2 M_\odot yr^{-1}$. On the other hand, with our current
selection we can reach SFR $\geq 10 M_\odot yr^{-1}$. Thus our selection allows to perform a
detailed study of the galaxy properties over a wide range of parameters,
such as stellar mass, age, SFR and dust content.

\section{The $IRX-\beta_{UV}$ relation at high redshift}

We have selected a sample of high redshift galaxies for which the whole rest-frame SED
from UV to near-IR is available, together with a robust proxy of the IR luminosity,
i.e., the amount of light absorbed by dust. The latter quantity could be directly used together
with the observed UV luminosity to derive the bolometric luminosity and hence the ongoing
SFR of the galaxy. However, in order to obtain other relevant physical parameters such
as the age and the mass of the galaxy, we need to go through the population synthesis
technique. This requires a few additional approximations concerning the dust attenuation,
the star formation law, the initial mass function and the metallicity of the stellar populations.

A detailed description of dust attenuation would require to exploit the chemical composition
and size of dust grains, the spatial distribution of dust and stars in the galaxy, and the
solutions of the radiative transfer equation (see Mathis 1990; Dwek 1998; Silva et al. 1998;
Draine 2003, and many others). However, many observational results can be rendered and
understood by treating the galaxy as a point-like source behind a screen of dust (see Calzetti
et al. 1994; Meurer et al. 1999).

To get insight on the form of the attenuation law, the observed UV data and the IR
luminosity can be combined to construct the so called $IRX-\beta_{UV}$ diagram; we recall that
$IRX$ is the IR to UV luminosity ratio and $\beta_{UV}$ is the rest-frame UV slope.
In spite of the fact that both such quantities are expected to be strongly affected
by the strength and by the wavelength dependence of the attenuation law,
Meurer et al. (1999) have pointed out the existence of a tight $IRX-\beta_{UV}$ correlation
in local starburst galaxies, indicating a quite universal form of the attenuation law.
Such correlation has been found to hold even in high
redshift $z\sim2$ samples (Reddy et al. 2006, 2010; Elbaz et al. 2011),
though with a larger scatter.

\begin{figure}
\centering
\includegraphics[width=12.0cm, angle=0]{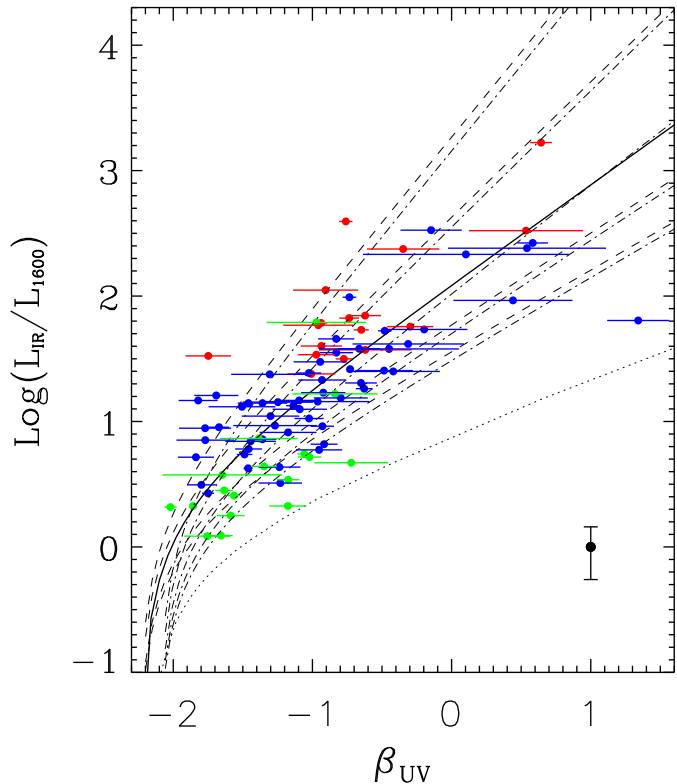}
\caption{The rest-frame $IRX-\beta_{UV}$ relation for our sample. Data points are
color-coded according to their IR luminosity  which is obtained from the rest-frame 8$\mu m$
luminosity by adopting a ratio $IR8 = 8.5$, see \S\ 2 for details.
Specifically, red points refer to $L_{IR}>10^{12}L_\odot$, blue to $10^{12}L_{\odot} \geq L_{IR} \geq 10^{11}L_\odot$, and green to $L_{IR}<10^{11}L_\odot$. The solid line shows the local Meurer relation, while the dotted line shows the corresponding relation obtained
on adopting the SMC extinction law. The dashed lines are our models with solar metallicity,
100$M_\odot yr^{-1}$ constant SFR,
and age of 0.1 Gyr for an increasing value of the $R_V$ parameter
in the Calzetti law (see Eq. [1]), from $R_V$ = 1.5 (bottom dashed curve) to $R_V$ = 10 (top
dashed curve). Along the curves, E(B-V) increases from 0 to a suitable upper limit. The
dot-dashed lines represent the same but with an average older age of 1 Gyr.
}
\end{figure}

For our sample, we have computed the UV slope by performing a linear fit to all available
photometric data with rest-frame central wavelengths falling in the range $1250\AA-2600\AA$.
The observed $L_{UV}$ is approximated by the quantity $L_{1600} = \lambda(1600)L_{\lambda}(1600)$,
while the \textit{observed} IR luminosity $L_{IR}$ is derived by the 8$\mu m$ rest-frame luminosity
by assuming the ratio $IR8=8.5$ as discussed in \S\ 2. The results are shown in Fig. 1
where galaxies are depicted with different colors, following their IR luminosities;
in particular, galaxies with $L_{IR} > 10^{12}L_\odot$ (ULIRGs) are in red, with
$10^{11} L_\odot \leq L_{IR} \leq 10^{12}L_\odot$ (LIRGs) are in
blue, with $L_{IR} < 10^{11}L_\odot$ (LLIRGs) are in green.
The solid line represents a model of solar
metallicity and a constant SFR = 100$M_\odot yr^{-1}$ with age of 0.1 Gyr. E(B-V)
increases along the curve, starting from 0 at the bottom left. We have adopted the Calzetti
law (cf. Eqs. [3] and [4] in Calzetti et al. 2000):
\begin{equation}
A(\lambda)=E(B-V)(f_\lambda+R_V)
\end{equation}
with $f_\lambda=2.659\times(-1.857+1.040/\lambda)$ for $0.63\mu m\leq \lambda \leq 2.20\mu m$
or $f_\lambda=2.659\times(-2.156+1.509/\lambda-0.198/\lambda^2+0.011/\lambda^3)$
for $0.12\mu m\leq \lambda \leq 0.63\mu m$, and $R_V=4.05$.

The observed luminosity of the model is given by
\begin{equation}
L_{obs}(\lambda)=L_{int}(\lambda)10^{-0.4A(\lambda)}
\end{equation}
where $L_{int}(\lambda)$ is the intrinsic stellar luminosity. The IR luminosity of the model
can be derived from
\begin{equation}
L_{IR}\approx L_{abs}=\int^{\infty}_{0}L_{int}(\lambda)[1-10^{-0.4A(\lambda)}]d\lambda ;
\end{equation}
where $L_{abs}$ is the absorbed luminosity. We recall that the Calzetti law reproduces
fairly well the Meurer relation for local starbursts.

From the Fig. 1 it is seen that on average our galaxies follow it consistently with other
recent studies at $z\sim2$ (Reddy et al. 2006, 2010; Elbaz et al. 2011). However, the individual
observations show a large scatter. In particular, the Meurer relation seems to be the lower
envelope for the more luminous galaxies with $L_{IR} > 10^{12} L_\odot$ and the upper envelope for
the less luminous ones with $L_{IR} < 10^{11} L_\odot$. The scatter is larger than
that allowed by the uncertainty in the estimate of IR luminosity through the $IR8$ ratio.
Indeed, systematically larger
$IRX$ ratios are also found in the local ULIRGs (Goldader et al. 2002), in $z\sim2$ ULIRGs
(Reddy et al. 2010) and in radio-selected submillimeter bright galaxies (Chapman et al.
2005). This bias is generally attributed to the fact that a significant fraction of the star
formation is strongly obscured at the UV band. The observed UV photons are not responsible for most of IR luminosity,
rather they are formed earlier and have already escaped from the dusty environments (Silva et al. 1998,
Reddy et al. 2006; Papovich et al. 2006; Chapman et al. 2005). On the other hand, values
of the $IRX$ ratios lower than predicted by the Meurer relation are generally ascribed to
different extinction laws. An example is that of the Small Magellanic Cloud (SMC), which is
represented by the dotted line (e.g. Pettini et al. 1998).

A few important points can be inferred from examining Fig. 1. First of all, the scatter
of the data points shows that the SFR cannot be robustly predicted on the basis of UV
SED alone. This highlights once more the importance of having a robust estimator of the
total IR luminosity. Second, we infer that it is impossible to fit simultaneously the
IR data on one side and the UV SED on the other, if one assumes
a single form (not strength) of the attenuation law (even by letting the age to change within
the values allowed by their observed redshift, see below). To clarify better this point we
have plotted in the same figure the results of other attenuation curves (dashed lines). They
have been obtained from the solid curve by varying the parameter $R_V$ from the Calzetti
value 4.05. It is seen that the data can be fairly well reproduced if we let $R_V$ to change
in the interval from 1.5 (bottom dashed curve) to 10 (top dashed curve). Increasing the
parameter $R_V$ flattens the attenuation law, as can be immediately seen by casting Eq. (1)
in the following form
\begin{equation}
A(\lambda)/A(V)=f_\lambda/R_V+1
\end{equation}
Such variation of the parameter $R_V$ is meant not only to render the effects of different
properties of the dust on the \emph{extinction}, but also to describe at first-order
other complex effects such as variations of the dust/star geometry.

Effects of age are also important, but only at low values of $\beta_{UV}\leq -2$ where the
$IRX-\beta_{UV}$ diagram becomes insensitive to $R_V$ because the attenuation is also small. On
the other hand, at high $\beta_{UV}\geq -2$ age effects are negligible,
as illustrated by the dot-dashed
lines that represent the same models as above but with an average older age (1 Gyr instead
of 0.1 Gyr).

\section{Stellar Population Synthesis Modeling}

In the light of the above discussion, now we turn to the issue of performing a robust
estimation of the galaxy parameters via the SED fitting technique. Our procedure consists of
modeling simultaneously both the detailed rest frame SED, from the far UV to the near-IR,
and the total luminosity absorbed by dust as predicted from the observed 24$\mu m$ flux. To
this purpose we use a set of simple stellar populations (SSPs) with solar metallicity and a
Chabrier (2003) IMF extending from 0.15 to 120$M_\odot$. The adopted SSPs are described in
more detail in the Appendix A.

As shown by Shapley et al. (2004, 2005), solar metallicity is a good approximation
in $z\geq2$ galaxies. In any case, we have checked that varying the chemical composition
from solar to $1/3$ solar does not change significantly our results. In fact, for younger stellar
populations the effect of metallicity is less severe than for older ones, as found in the local
Universe. Different kinds of IMFs such as Salpeter (1955) or Kroupa (2001), and different
stellar mass ranges of them will change the final stellar mass up to a factor $\leq1.6$.

For the sake of simplicity, in this paper we have considered the case of constant SFR.
Other parameterizations are also quite used in the literature, involving an exponential
increase, exponential decrease, or sequence of both. In sophisticated galaxy formation
models (e.g., Granato et al. 2004; Fan et al. 2008, 2010; Lapi et al. 2006, 2011),
the SFR initially increases, attains
a closely constant value, and then is quenched by the energetic feedback from supernovae
or quasars (soon after one Gyr for strongly star-forming sources, and after several Gyrs for weakly
star-forming ones). Then the constant SFR adopted here is a fairly good representation
of the average behavior expected for these objects during most of the time they spend as
star-forming sources. This is also in agreement with the recent analysis by Reddy
et al. (2012b) of a large spectroscopic galaxy sample at redshift $1.4\leq z \leq 3.7$ based on
\emph{Hubble} and \emph{Spitzer} observations.

We include in the calculation the total IR luminosity, and treat the predicted value
as a constraint in the SED-fitting minimization procedure. The predicted IR luminosity is
computed after Eq. (3), while the observed IR luminosity is derived from the 24$\mu m$ flux
as described in \S\ 2. The attenuation law is assumed to be independent of the age and
parameterized by means of Eq. (1). However, according to the previous discussion on the
$IRX-\beta_{UV}$ relation in \S\ 3, the $R_V$ value in Eq. (1) is considered
as a free parameter in the
fitting procedure (see also the discussion in Calzetti et al. 2000). It was let to vary
in the range $1.5-10$, which as discussed before provides a fair representation of the dispersion in Fig. 1.

The model contains four parameters: the strength of the constant SFR, the age of the
galaxy $t_g$ (limited only by the consistency with the observed redshift), the strength of the
attenuation parameterized by E(B-V), and the parameter $R_V$ . Then we
compute the intrinsic spectrum of the galaxy, to which we apply the internal reddening. The
absorbed spectrum is integrated to provide the model IR luminosity which is constrained
by the corresponding observed quantity. Finally, after applying the attenuation due to the
intervening intergalactic neutral hydrogen (Madau 1995), we convolve the spectrum with
the transmission filters, and obtain the model fluxes to be compared with the observed ones.
The best fit model is obtained by minimization of the merit function
\begin{equation}\label{eq_MF}
{\rm MF}=\sum_{i=1}^{n}\left(\frac{M_i-O_i}{E_i}\right)^2
\end{equation}
where $M_i$, $O_i$ and $E_i$ are the model values, the observed values
and the observational errors,
including the IR luminosity. The minimization is performed with an Adaptive Simulated
Annealing algorithm (Ingber 1989).

\begin{figure}
\centering
\includegraphics[width=0.3\textwidth,angle=90]{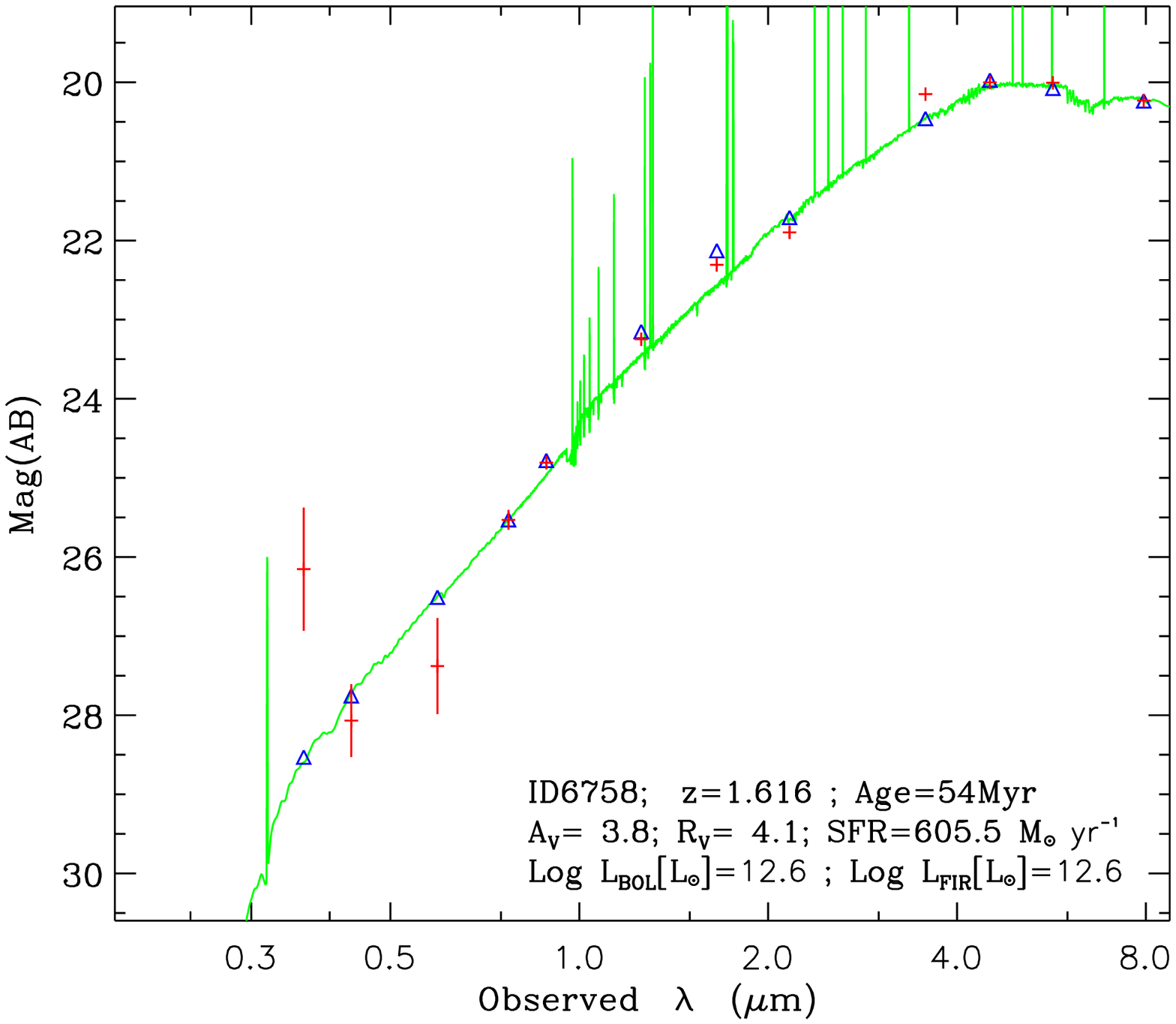}
\includegraphics[width=0.3\textwidth,angle=90]{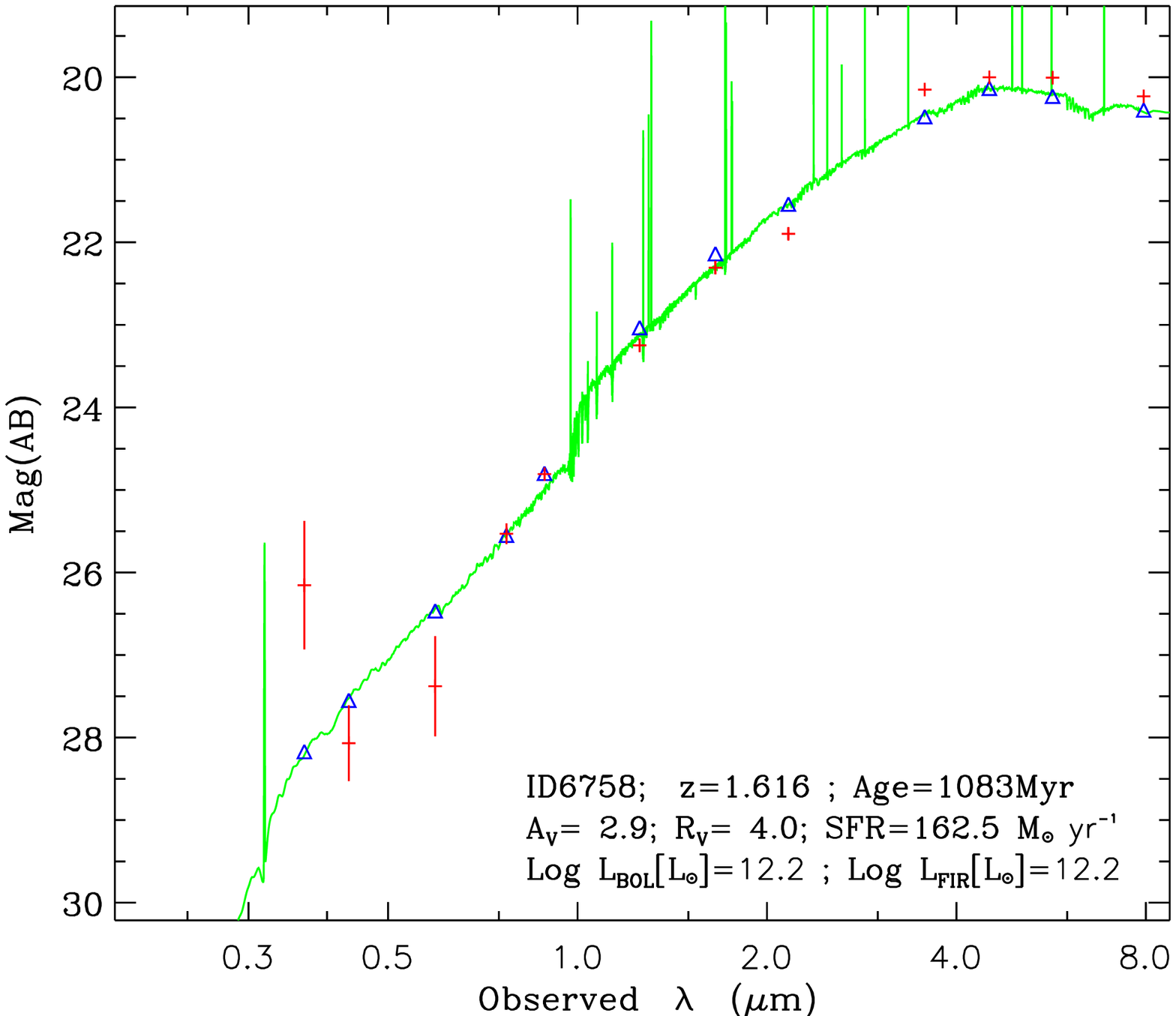}
\includegraphics[width=0.3\textwidth,angle=90]{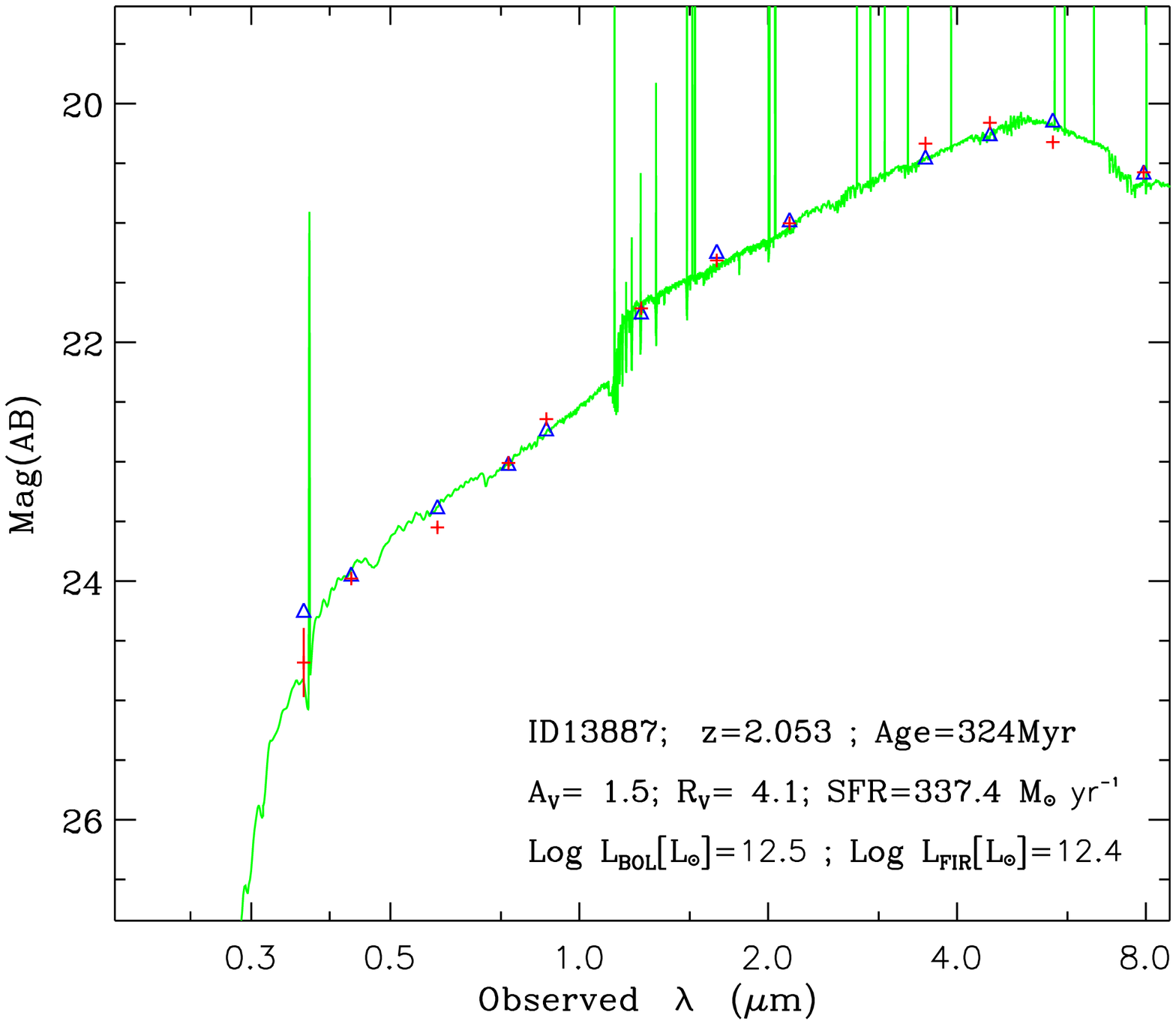}
\includegraphics[width=0.3\textwidth,angle=90]{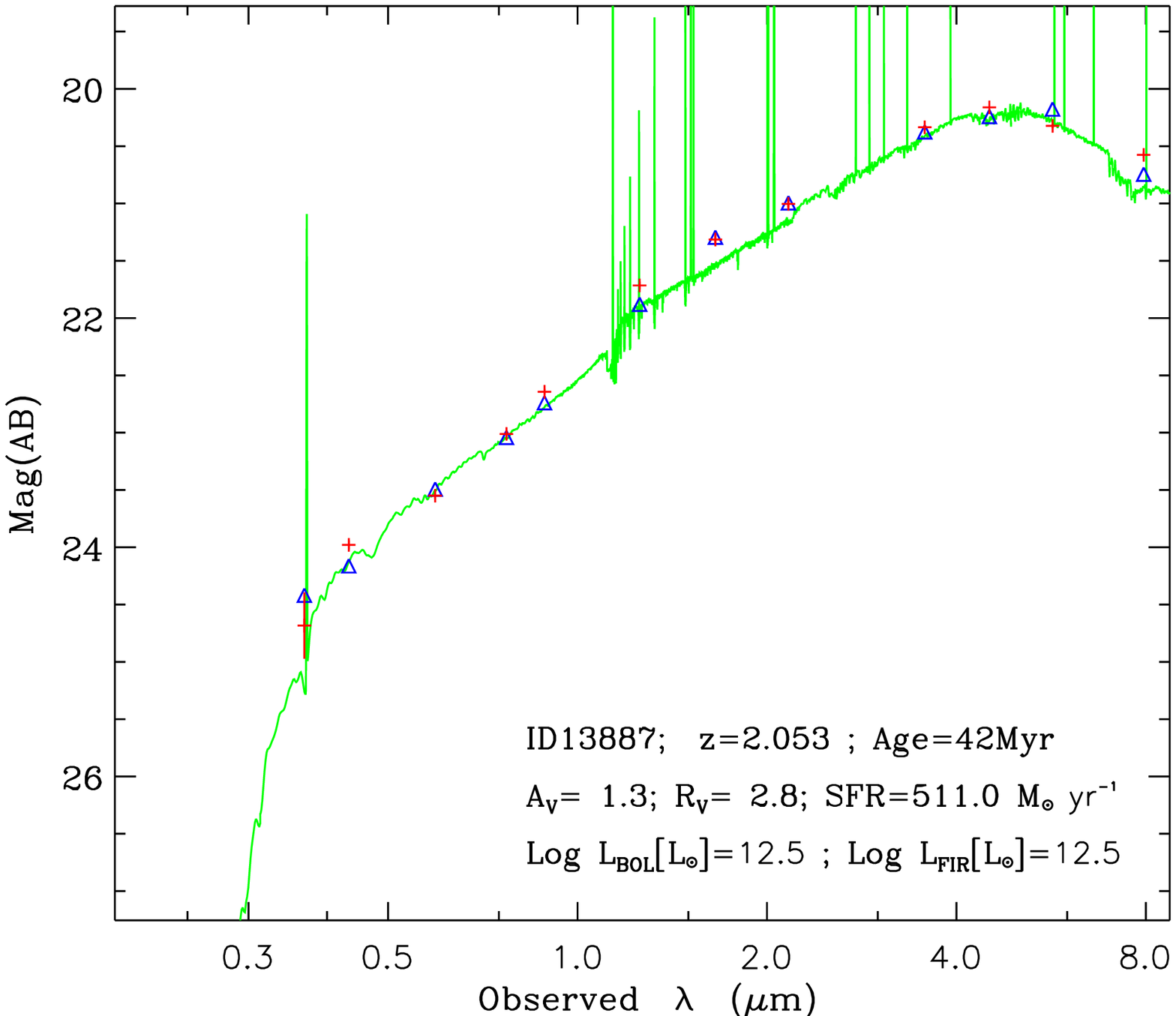}
\caption{Best fit models obtained with (right panels) or without (left panels) the IR
prior derived from the $IR8$ relation. In the former case the $R_V$ parameter of the attenuation
law is not fixed while in the latter case the value $R_V = 4.05$ is adopted. The observed
fluxes are plotted with crosses while corresponding model fluxes, after convolution with the
transmission filters, are plotted with triangles. The solid curve is the best fit model. Both
the SFR and the attenuation are kept constant with the age of the stellar populations. Error
bars on fluxes are also plotted but are in general smaller than the symbol size.}
\end{figure}

We stress that the IR luminosity is a prior that provides the unbiased level of the
SFR. For a given IMF, the error on the SFR is mainly observational, though there is some
dependence on the adopted attenuation law and on the metallicity of the stellar populations, as already
discussed by Shapley et al. (2005). Determining the correct shape of the SFR will require
additional information, such as spectral emission, absorption lines, and (non-)thermal radio
continuum. These effects probe the diverse contributions from stellar populations of different
ages and can be used, in principle, to disentangle the correct shape of the SFR (e.g. Bressan
et al. 2002; Vega et al. 2008).

With the SFR essentially determined by the IR luminosity, the observed UV continuum
mainly sets the level of the attenuation. This is also a quite robust result because
the IR luminosity is mainly produced by the absorbtion of the UV light. The observed
SED from the UV to the near-IR results from the combination of the attenuation shape
and of the galaxy age. There can be some level of degeneracy between age, metallicity and
attenuation shape, but the number of constraints is also quite large. Finally, the stellar mass
$M_\star$ is obtained by multiplying the SFR by the galaxy age $t_g$ which amounts to the age of
the oldest stellar population required by the fitting procedure.

\section{Results}

In this Section we present and discuss our main findings.

\subsection{Comparison between the two different synthesis methods}

\begin{figure}
\centering
\includegraphics[width=12.0cm, angle=0]{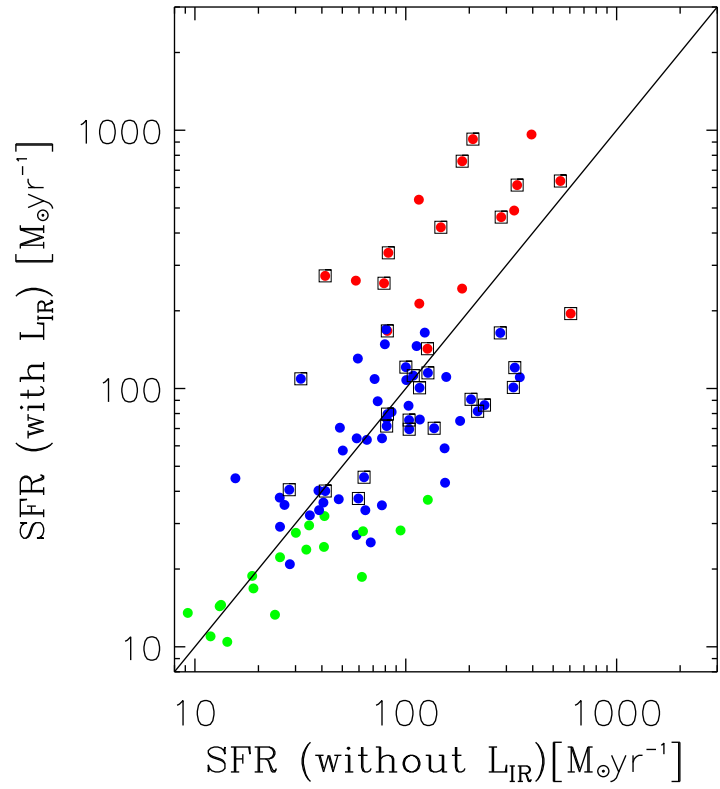}
\caption{Comparison of the SED-inferred SFR obtained with and without the IR prior.
Color-code is the same as in Fig. 1; objects with error $\leq$15\% on the 24 $\mu m$ flux have been surrounded
by boxes.
}
\end{figure}

To begin with, we compare the results obtained with and without the use of the IR
prior. We remind that in the latter case, which we name the standard procedure, we have
also assumed a definite attenuation shape, i.e., a Calzetti law with $R_V = 4.05$. Fig. 2
(right panels) shows a few examples of best fit models obtained with the IR prior.
The red crosses refer to the observed fluxes, while the triangles to the corresponding model
fluxes (after convolution with the transmission filters).  All the galaxies have also been
fitted with the standard procedure and the best fit models are shown in Fig. 2 (left panels).
We see that even without the IR prior the best fit is in general fairly good , however
the inferred star formation is different.

In Fig. 3 we compare the SFR derived with and without the IR prior for the whole
sample; symbols and color-codes are the same as in Fig. 1 for ULIRGs, LIRGs and LLIRGs.
Data with uncertainties $\leq15\%$ on the 24 $\mu m$ flux are surrounded by boxes. For almost
all the ULIRGs the SFR obtained without the IR prior is underestimated, even by a large
factor. On the other hand, LIRGs distribute across the one-tone relation with a quite
large dispersion. The latter decreases for the less luminous LLIRGs. Part of the dispersion is
certainly due to our assumed value $IR8\approx8.5$, since less compact star-forming galaxies show
a lower average value ($IR8\approx5$) while more compact star-forming objects show a higher
average value ($IR8\approx13$). Thus we could have overestimated the IR luminosity of the less
compact galaxies and underestimated that of more compact objects, by about 50\%, which
translates into a similar scatter in the plot of Fig. 2. However, the dispersion is actually
significantly larger than this estimate, reaching in some cases a factor of 5. Therefore it must
be intrinsic to some extent.

\begin{figure}
\centering
\includegraphics[width=12.0cm, angle=0]{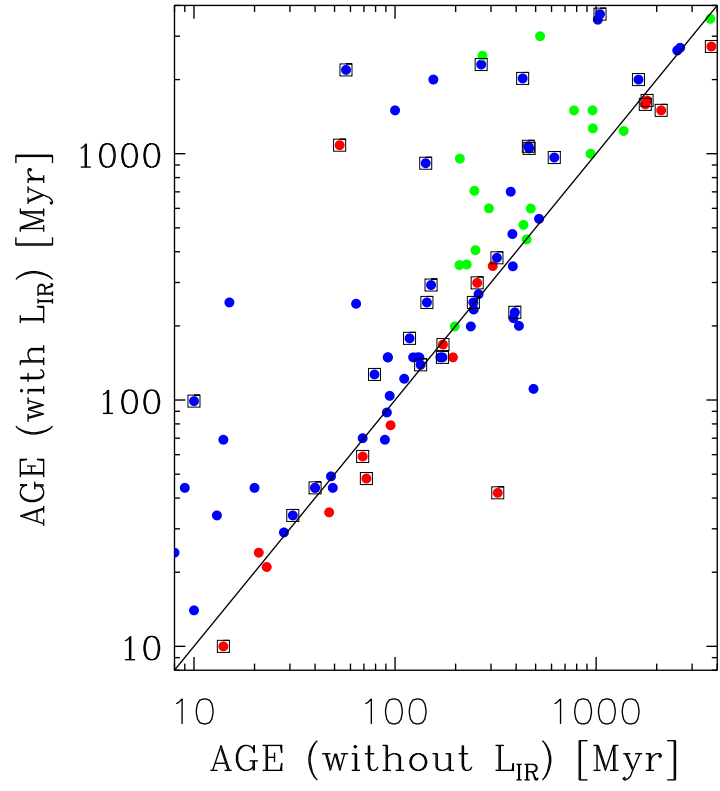}
\caption{Same as Fig. 3 for the age.}
\end{figure}

The presence of galaxies for which the predicted SFR is underestimated is actually
expected, because a significant fraction of the SFR could be entirely hidden due to strong
attenuation. It has already been shown that in several local LIRGs and ULIRGs the SFR
determined by the $H_\alpha$ or $Pa_\alpha$ luminosity, even corrected for attenuation
with the Balmer
decrement method, turns out to be lower by even a factor of three than that obtained from
the IR luminosity (see Poggianti et al. 2001; Vald\'es et al. 2005). On the other hand, the
presence of objects for which the standard SED fitting technique overestimates the IR
flux is a bit more intriguing. As a matter of fact, in these objects the reddening deduced
from the UV/near-IR SED, if interpreted by means of the usual Calzetti law, would require
a higher overall attenuation and this would produce a IR luminosity which is larger than
the observed one. For the LLIRGs the use of a prior in the IR luminosity is not as important
as for the higher luminosity galaxies. We will come back to this point at the end of the
Section.

To proceed with the comparison between the two synthesis methods, we show in Fig. 4
the galaxy ages and in Fig. 5 the stellar masses. The latter mass is computed as the integral
of the SFR over the galaxy age, and it does not take into account stellar recycling into the
interstellar medium. We see that the age determination is, in general, quite independent from
the IR prior. However, there is a minority of galaxies for which there is a large discrepancy
between the two age determinations, with the age determined with the IR prior turning out
to be significantly larger than those based on the standard procedure. Furthermore, when
the IR prior implies larger ages then it implies also lower SFRs, so that the differences on
the masses obtained with the two methods are smoothed out. Nevertheless, it is also clear
that the standard procedure underestimates the total mass by an average factor of about
50\%, with some individual values that may be offset by even a factor of 4.

\begin{figure}
\centering
\includegraphics[width=12.0cm, angle=0]{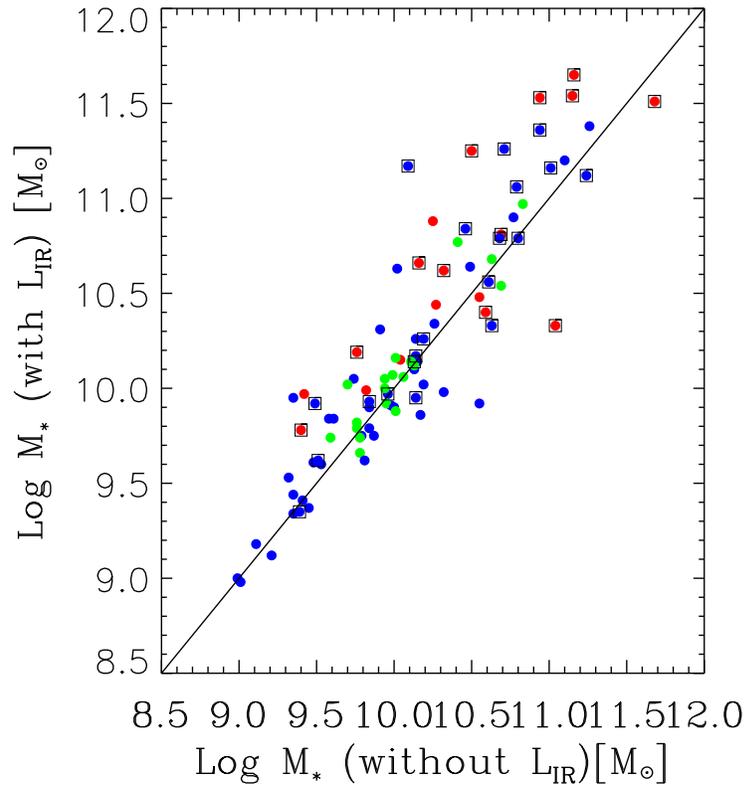}
\caption{Same as Fig. 3 for the stellar masses.}
\end{figure}

For LLIRGs the masses derived using the two different methods are in fair agreement. In
this class the attenuation is not strong enough to totally enshroud the star-forming regions
and the SFR can be obtained directly from the fit of the UV/near-IR SED. Actually, the
SFRs obtained with the standard procedure are even slightly overestimated for the majority
of these sources. The reason behind this effect is that the standard procedure assumes the
Calzetti law, whose $R_V$ is larger than the characteristic value of low star-forming objects, as
can be seen from Fig. 1. In fact, the median value of $R_V$ for objects in the LLIRG class is
$R_V\approx3.0$. Notice that, as shown in the following, the average age of these LLIRGs is about
0.9 Gyr, and thus their location in Fig. 1 must be compared with the dashed line referring
to a model with $R_V = 3$ and an age of 1 Gyr. Once repeated with such a lower value of $R_V$ ,
the results obtained with the standard procedure compare fairly well to those obtained with
the IR prior.

This point is interesting because it allows us to increase the sample for which a robust
estimate of the parameters can be obtained with SED fitting technique. In particular the
number of LLIRGs is small and to draw more robust conclusions it would be desirable to
extend the above analysis to the full $\mz$-selected sample. We remind that, in order to adopt the
IR prior, we use only the sources detected at 24$\mu m$, with a detection threshold corresponding
to a SFR of about $10M_\odot yr^{-1}$. With this selection criteria we cut about half of our original
$\mz$-selected sample, with the majority of the excluded objects
belonging to the lower luminosity
class. The experiment described above shows that we may release the constraint of the
IR prior in the subsample of galaxies undetected at 24$\mu m$ provided that, in the standard
procedure, we adopt a Calzetti law with a lower value of $R_V\approx 3.0$.

We have thus repeated the standard procedure with a lower value of $R_V$ for the 24$\mu m$
undetected sources. The SED fitting for these galaxies are not as good as those for the
sample detected at 24$\mu m$. These galaxies are intrinsically less luminous and the photometric
uncertainties are on average larger, being typically between 0.2-0.5 mag in about half of
the bands. In several objects the derived SFR is higher than the average threshold imposed
by the 24$\mu m$ criterion, but the object was not detected at 24$\mu m$ either because of a lower
exposure depth or because the SED-inferred SFR was significantly affected by photometric
errors. We thus excluded from this sample those objects with a derived
SFR$\geq 30M_\odot yr^{-1}$ for
which, according to Fig. 3, the knowledge of the IR prior is mandatory. After this further cut
we remain with 41 objects with $L_{IR}\leq 10^{11} L_{\odot}$.
This subsample is constituted by the 24 $\mu m$
undetected galaxies for which the standard procedure allows a fairly robust estimate of the
physical parameters. In the following we will consider also the galaxies of this subsample
that we will refer to as 24$\mu m$ undetected LLIRGs.

\subsection{SFR and Attenuation}

\begin{figure}
\centering
\includegraphics[width=12.0cm, angle=0]{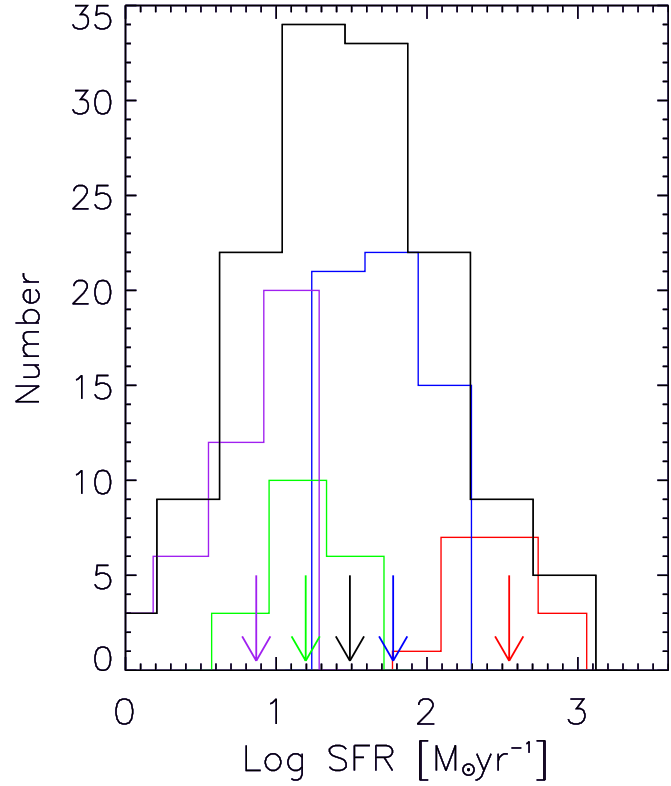}
\caption{Distribution of the SFR in our galaxy sample. Histograms highlight the distributions
and arrows mark their average values. Color-code refers to the subsamples defined
in Fig. 1 (red for ULIRGs, blue for LIRGs, and green for LLIRGs), to the 24 $\mu m$ undetected
LLIRGs (purple) and to the overall sample (black).}
\end{figure}

Fig. 6 shows the distribution of the SFRs for our sample. The primary selection in the
$\mz$-band allows us to detect galaxies with a wide range of SFRs,
from a few to $\geq 10^3 M_\odot yr^{-1}$.
We recall from the above that in the SED fitting technique we have assumed $R_V$ as a free
parameter. The upper left panel in Fig. 7 shows the resulting values of $R_V$ as a function of
the SFR. We notice a broad tendency of $R_V$ and the associated scatter to increase with the
strength of the SFR. Part of the scatter could be due to the error accumulated in the process
of obtaining the IR luminosity, whose largest contribution is given by the assumption of
a single average relation with the 8$\mu m$ luminosity $IR8$. However, a similar scatter is also
evident in analogous plots for local sources, for which this uncertainty is smaller, suggesting
that there is an intrinsic dispersion connected with the shape of the attenuation law.

As already discussed in \S\ 3, at decreasing $\beta_{UV}$ the loci
with different $R_V$ converge to
a single curve. Therefore one would expect that the scatter in the inferred $R_V$ , due to
errors in either $\beta_{UV}$ or $L_{IR}$, increases at decreasing $\beta_{UV}$ .
On the other hand, all LLIRGs with
small $\beta_{UV}$ , exhibit a quite narrow scatter around $R_V\approx 3$.
Note that the peculiar object
represented by the green dot appearing in the upper left corner features a very high
value $R_V = 10$ and very old age 3 Gyr. It has been retained in the sample because formally
the $\chi^2$ value of the SED fitting is low, but a visual inspection shows the corresponding data
to suffer large uncertainties.

To better stress the meaning of the $R_V$ parameter, we remind that for the Calzetti law
$R_V\approx4$ and that, by increasing $R_V$ ,
the attenuation law becomes progressively flatter, i.e.,
more neutral. In particular, while for a typical Calzetti law the ratio of the attenuation in
the K and V band is $A_K/A_V \approx 0.1$, for $R_V = 10$
it increases to $A_K/A_V\approx0.6$. Thus we
expect that there will be a number of galaxies for which the attenuation in the near-IR band
is not negligible with respect to that in the visual band, as usually assumed in star-forming
galaxies. Indeed, several ULIRGs of our sample are well fitted by an attenuation law which
is more neutral than the typical Calzetti law. In order to reproduce their observed IR
luminosity together with the observed UV slope, a flat attenuation law is required, otherwise
the predicted spectral slope would be too steep. This is consistent with the evidence that,
in several local ULIRGs, the molecular clouds associated with the star-forming regions are
optically thick even in the near-IR (Vega et al. 2008). In such galaxies the bulk of IR and UV
luminosity could originate from different regions, as found in
nearby starburst galaxies (Goldader et al. 2002). Thus the emerging UV radiation could not
even directly mirror the global attenuation of the star-formation process, because the former
may originate in regions that are occasionally located in a less dusty ambient, or in a few
regions where young stars emerge from the dusty ambient in a relatively shorter timescale
(Silva et al. 1998; Panuzzo et al 2003). Our parametrization is meant to be a fair description
of the attenuation resulting from the combination of all the intervening processes such as
age-dependent extinction, geometrical effects, and real differences in the dust mixture.

\begin{figure}
\centering
\includegraphics[width=0.4\textwidth,angle=0]{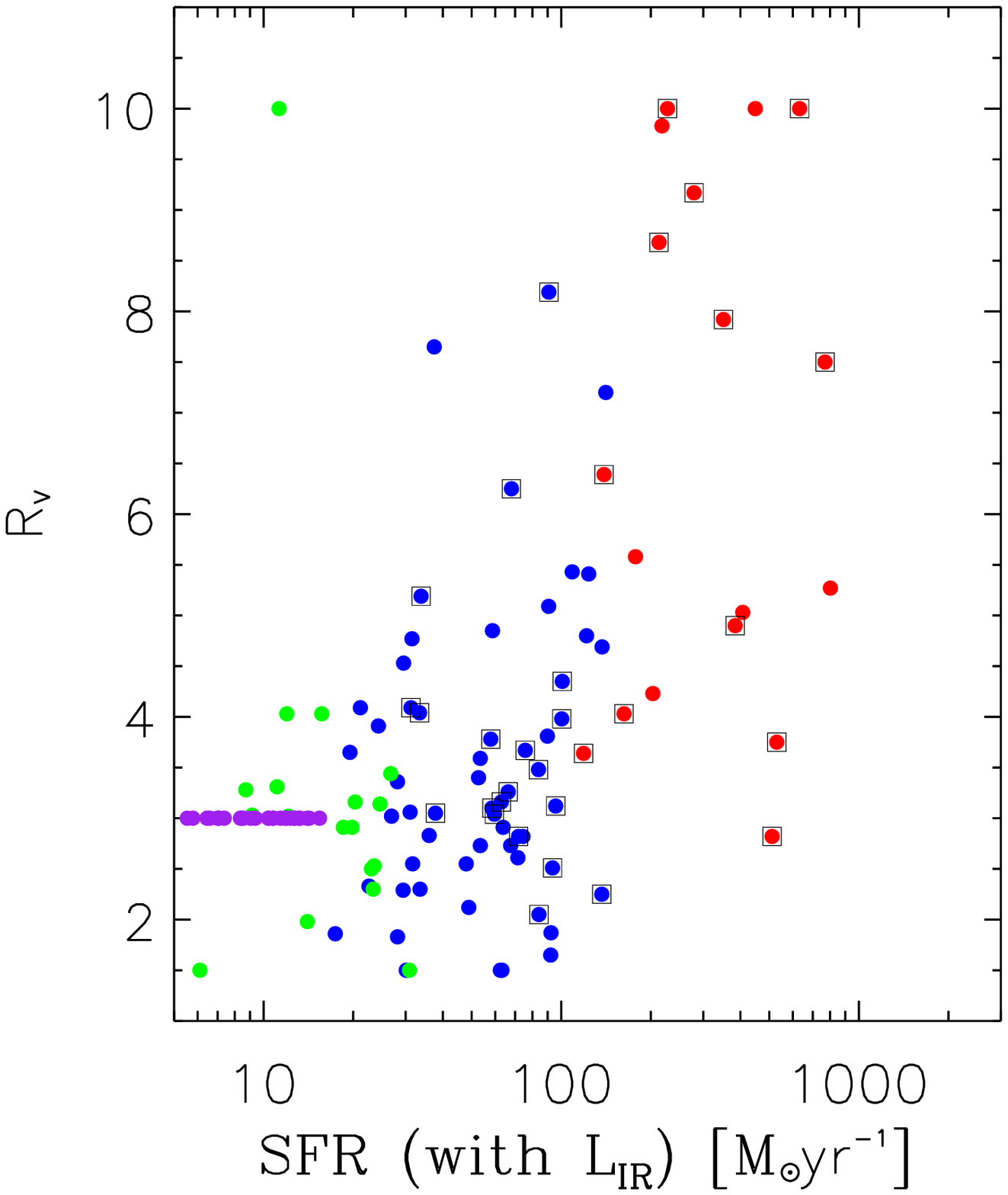}
\includegraphics[width=0.4\textwidth,angle=0]{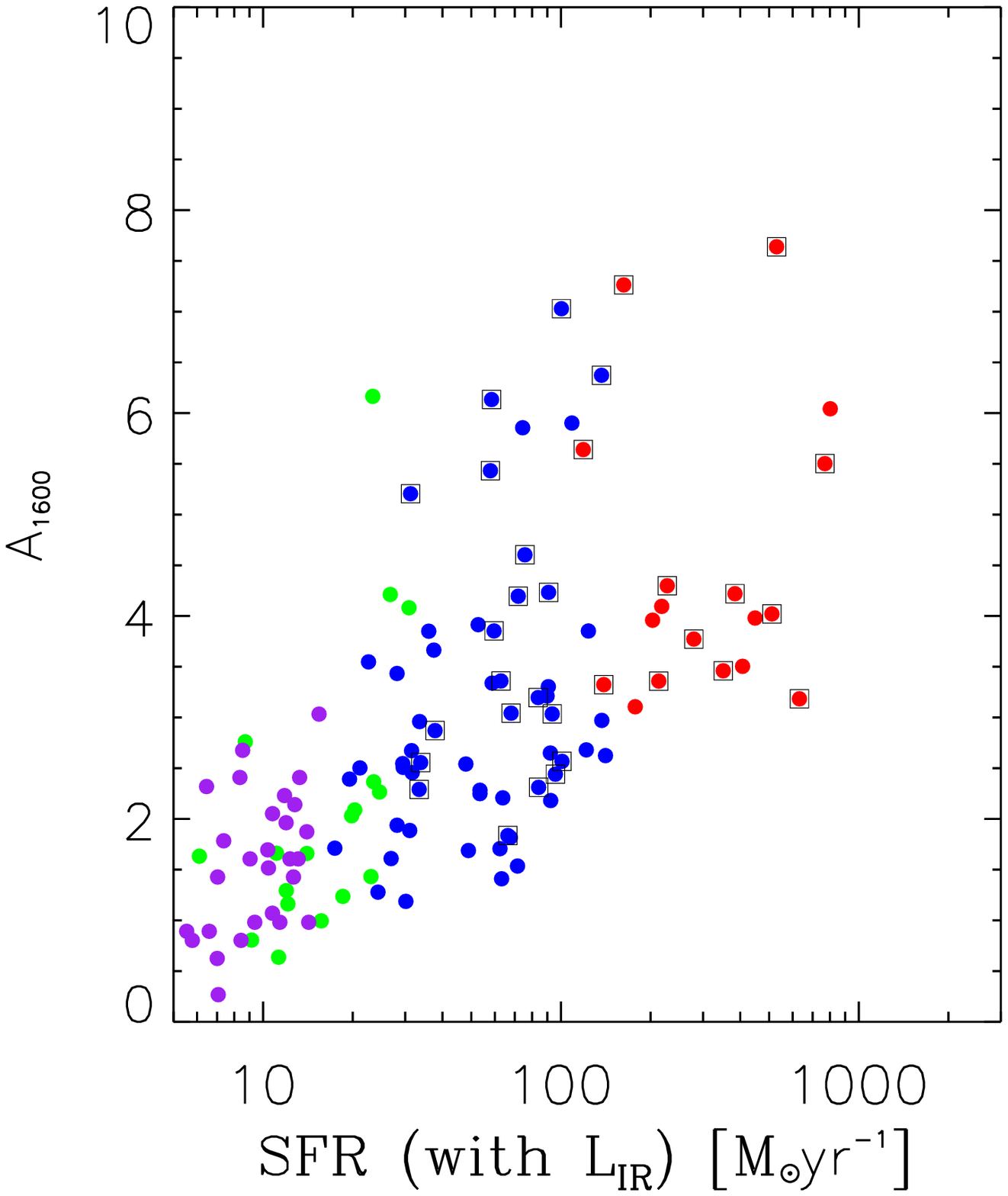}
\includegraphics[width=0.4\textwidth,angle=0]{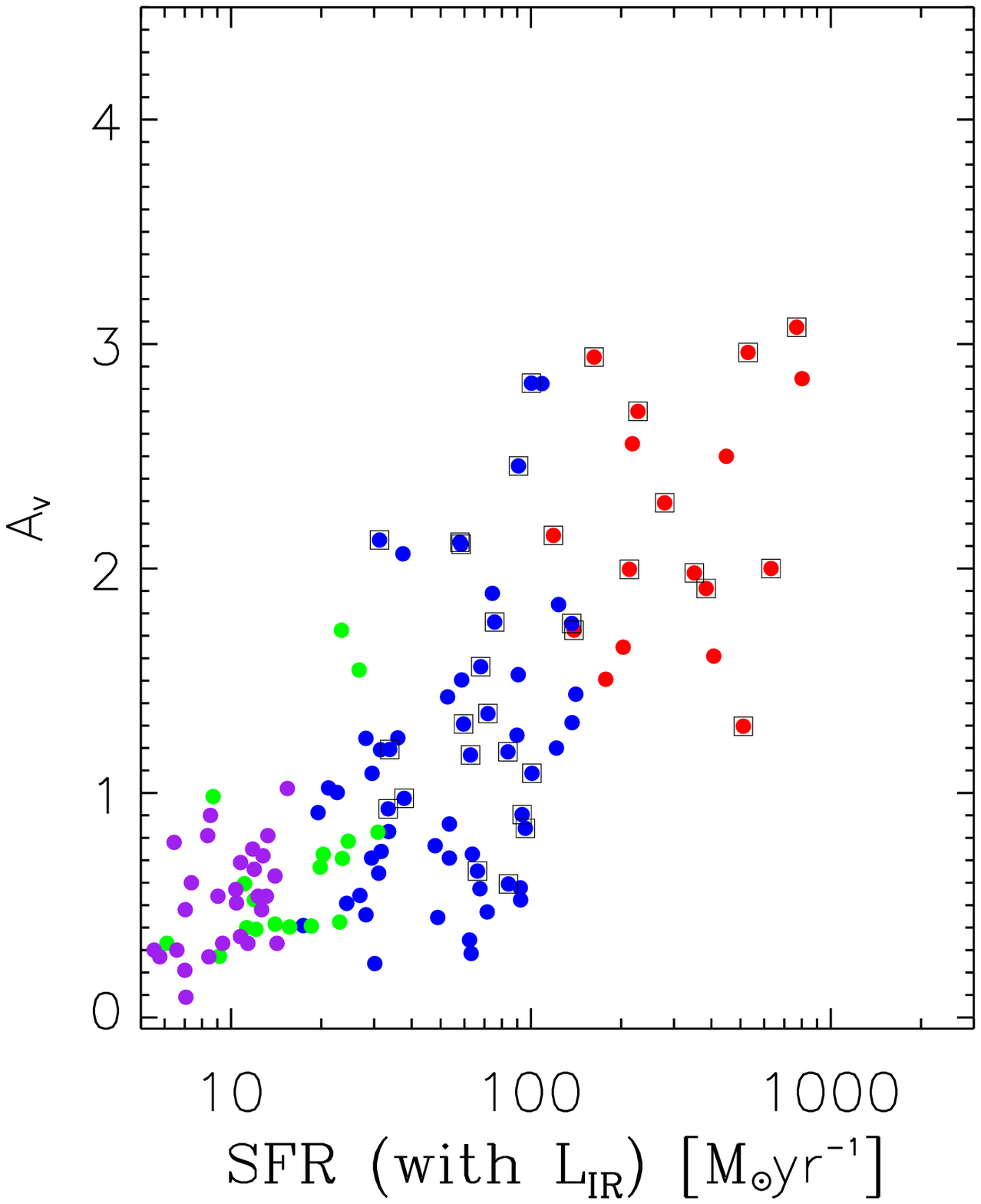}
\includegraphics[width=0.4\textwidth,angle=0]{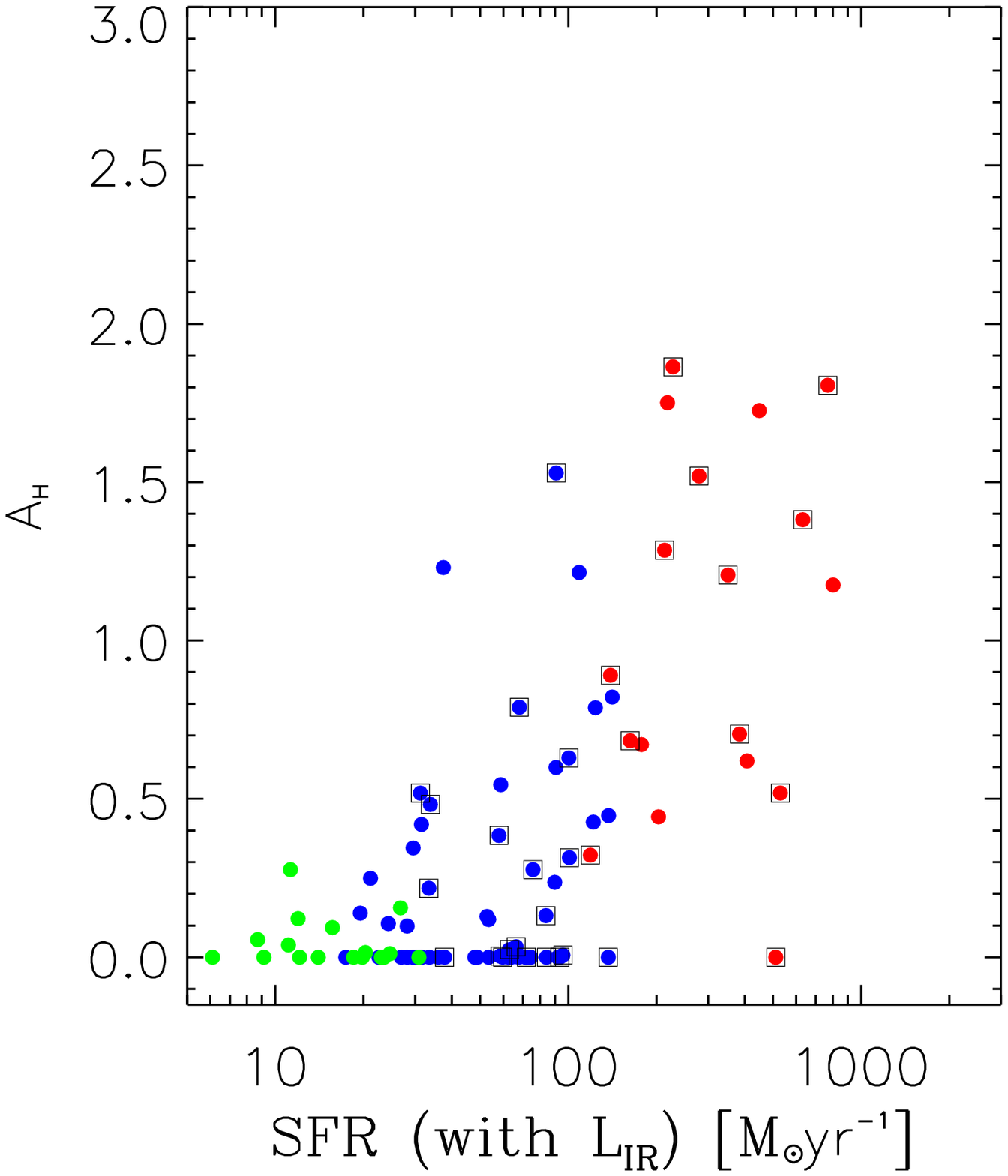}
\caption{Dependence of the selective to global attenuation ($R_V$ ), and of the attenuation in
the UV ($A_{1600}$), in the visual ($A_V$ ) and in the H-band ($A_H$) on the SFR. Color-code is the same as in
Fig. 6.}
\end{figure}

In the other panels of Fig. 7 we show the relation between the SFR and the attenuation
in the UV ($A_{UV}$ ), in the optical ($A_V$ )
and in the near-IR ($A_H$). In all cases the attenuation
grows with the SFR, as found by other authors, though the scatter is large. The scatter of
the attenuation in the UV is not unexpected and is even consistent with the knotty nature
of UV images. The distribution of the UV attenuation is shown in Fig. 8. The median values
for the different subclasses decrease at decreasing luminosity, while for the whole sample
the median value is $A_{UV}\approx2.3$ mag and the maximum value is $A_{UV}\approx9$ mag.
However, the distribution is clearly asymmetrical with a sharp drop
above $A_{UV}\approx4$ mag. This could be
related to the strong UV dimming suffered by galaxies with large $A_{UV}$ . Indeed, two thirds
of the galaxies above the sharp drop are ULIRGs. As to the near-IR we notice that the
attenuation there may reach values that are significantly higher than those expected in the
UV (or in the visual) by assuming a standard extinction law.

\begin{figure}
\centering
\includegraphics[width=12.0cm, angle=0]{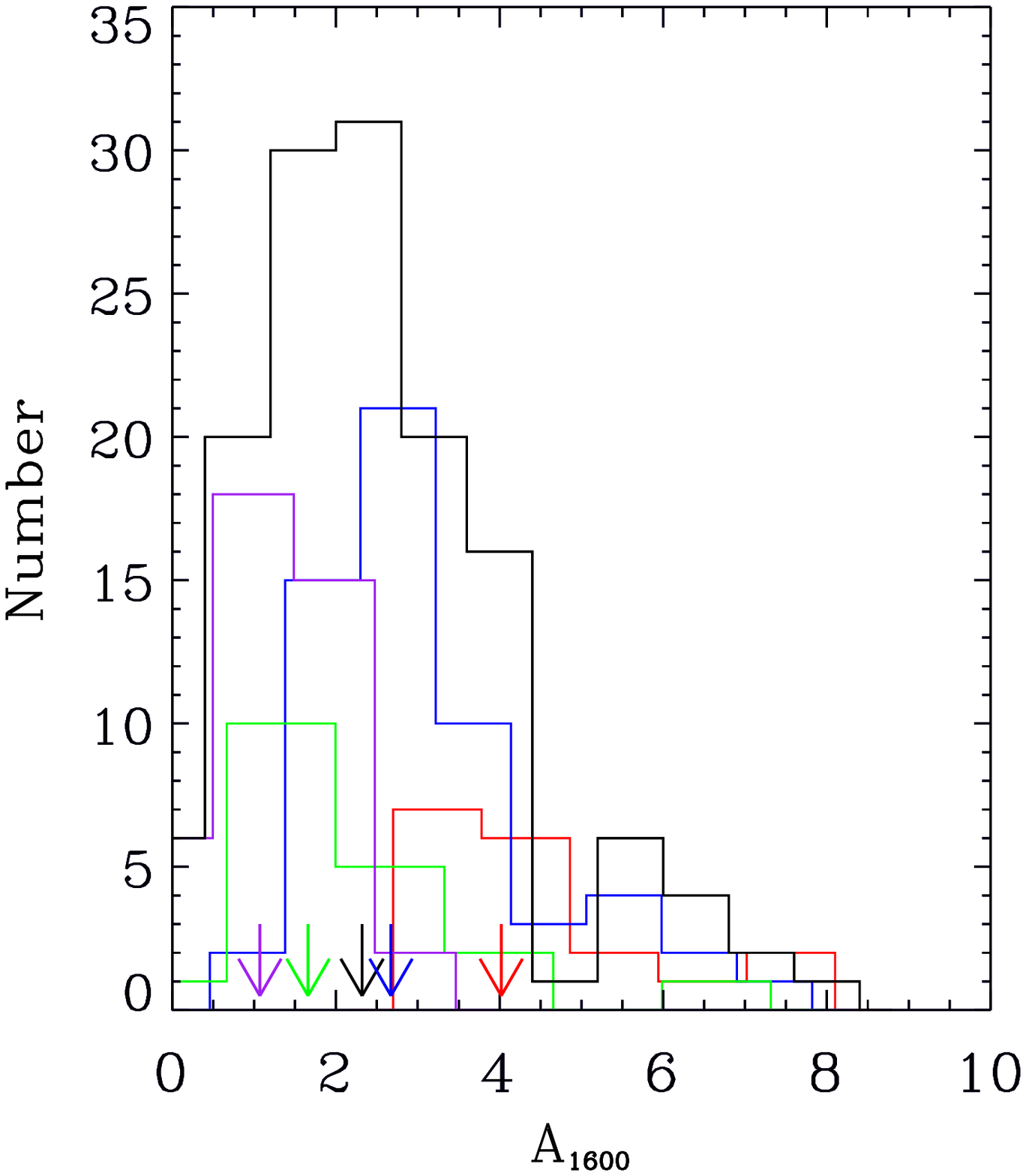}
\caption{Distribution of the UV attenuation ($A_{1600}$) in our galaxy sample. Color-code
and line styles are the same as in Fig. 6.}
\end{figure}

\subsection{Stellar Masses and Ages}

\begin{figure}
\centering
\includegraphics[width=12.0cm, angle=0]{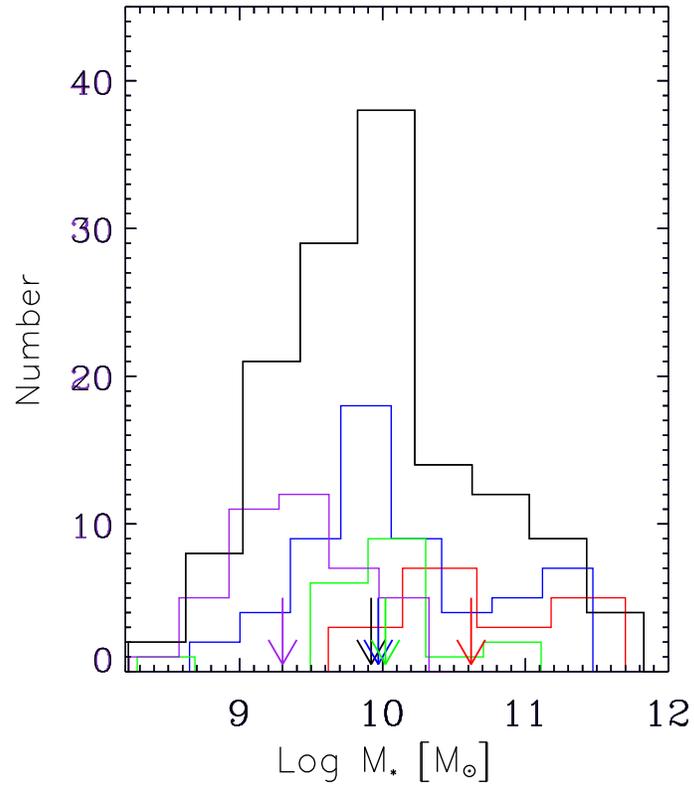}
\caption{Distribution of the stellar masses in our galaxy sample. Color-code and linestyles are the same as in
Fig. 6.}
\end{figure}

\begin{figure}
\centering
\includegraphics[width=12.0cm, angle=0]{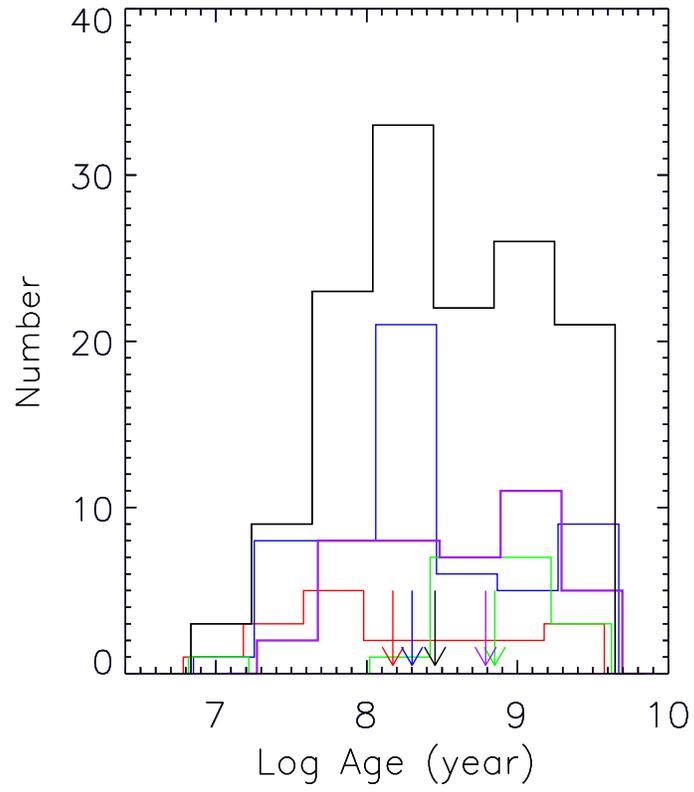}
\caption{Age distribution of our galaxy sample. Color-code and linestyles are the same as in Fig. 6.}
\end{figure}

\begin{figure}
\centering
\includegraphics[width=0.6\textwidth, angle=90]{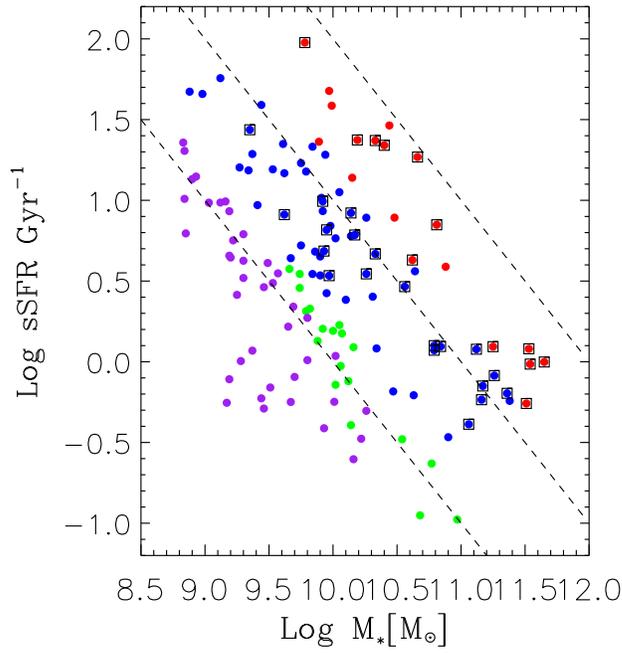}
\caption{Specific star formation rate as a function of the stellar mass. Color-code is the same as in
Fig. 6. Dashed lines refer to a constant SFR of 10, $10^2$ and
$10^3M_\odot yr^{-1}$ from left to right, respectively.}
\end{figure}

\begin{figure}
\centering
\includegraphics[width=0.4\textwidth,angle=0]{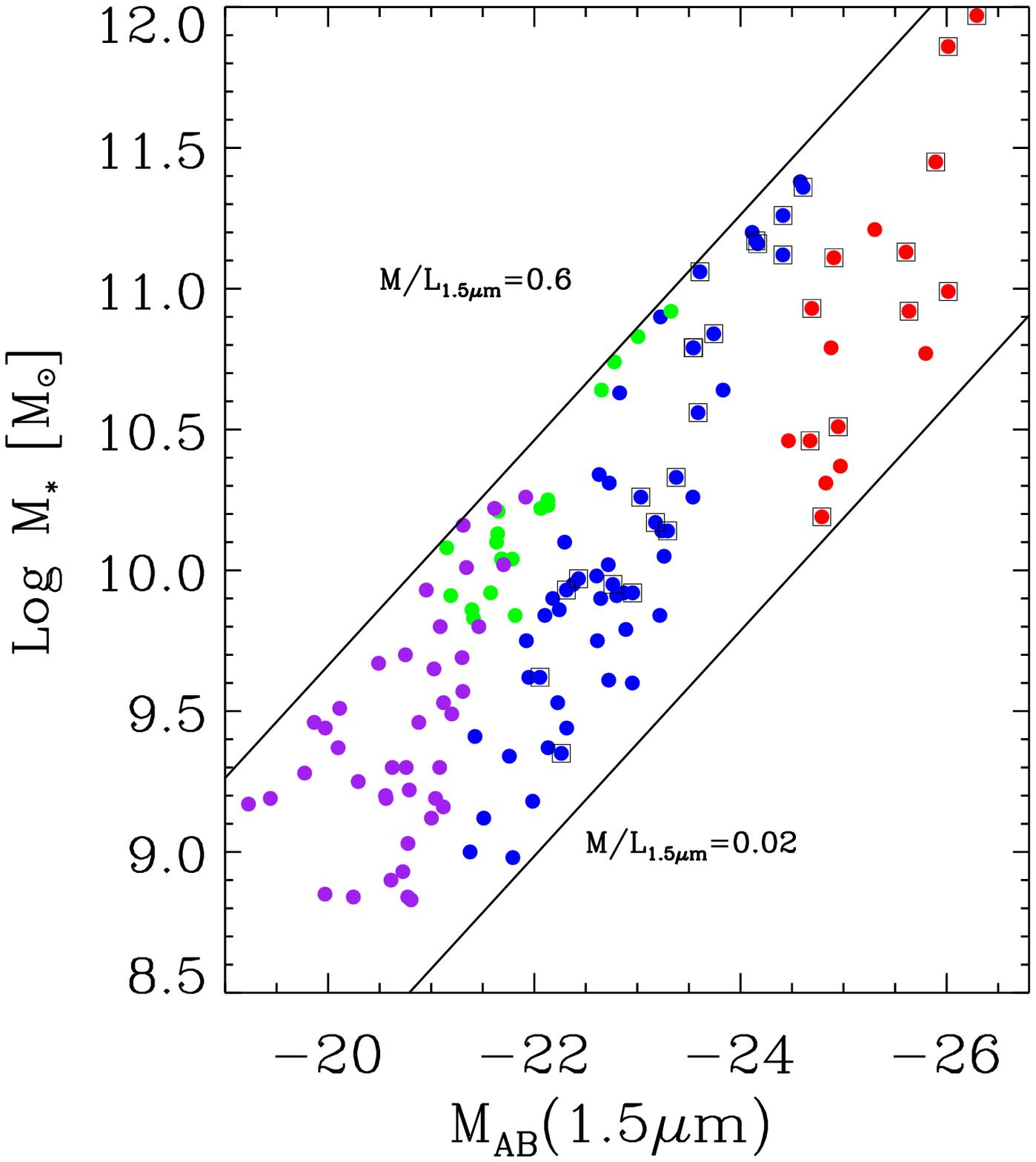}
\includegraphics[width=0.4\textwidth,angle=0]{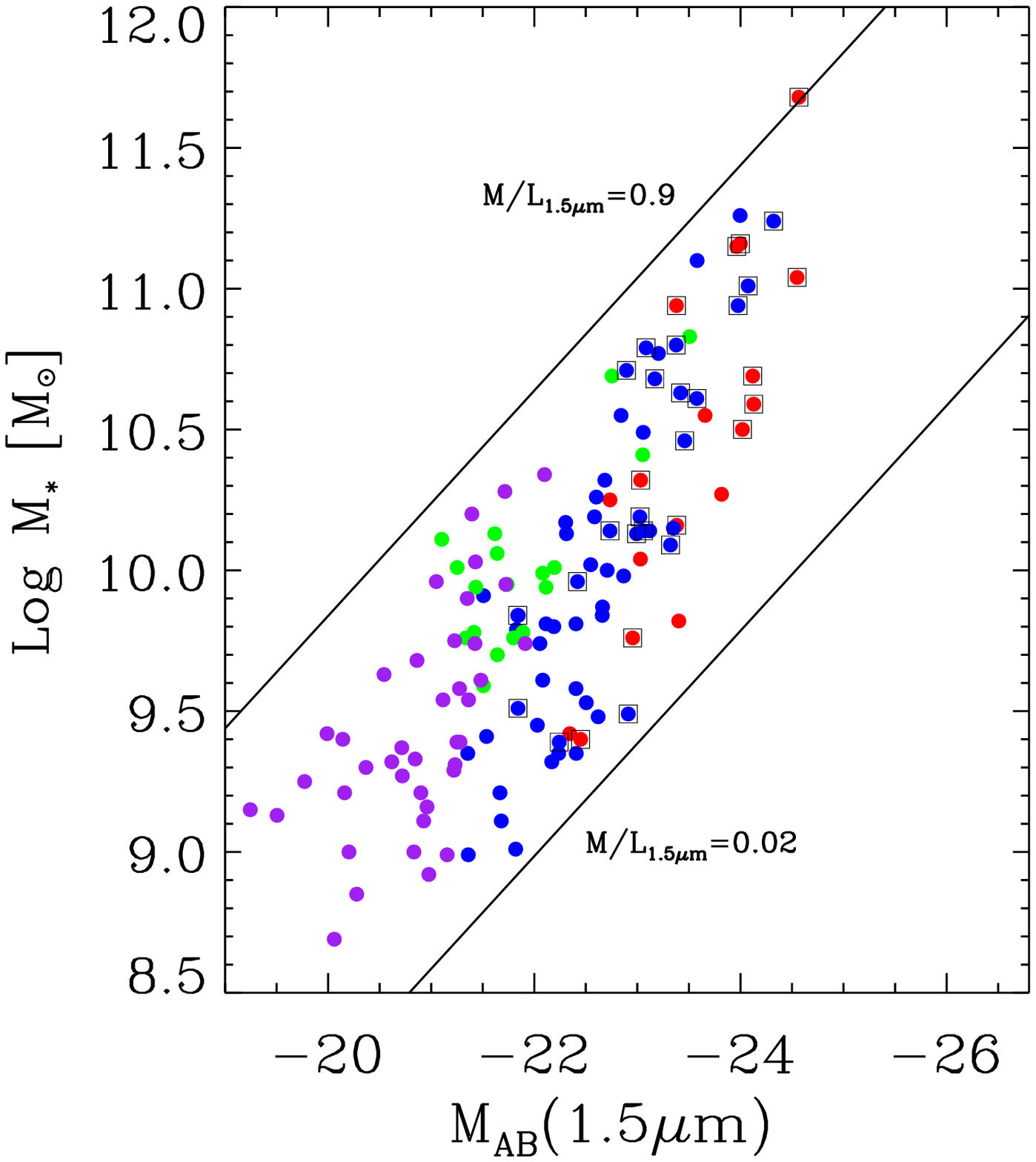}
\caption{Left Panel: intrinsic mass to light ratio of our galaxies in the rest-frame H-band,
in solar masses per solar H-band luminosity, vs. the absolute H-band magnitude. Right
panel: the corresponding diagram obtained with the standard procedure. Color-code is the same as in
Fig. 6.}
\end{figure}

The stellar mass distribution of our sample is shown in Fig. 9. The masses range
from $M_\star\approx 10^9M_\odot$ to
$M_\star\approx 4\times10^{11}M_\odot$. ULIRGs have a quite flat mass distribution up
to the higher values, and a median value of $M_\star\approx 4.2\times 10^{10}M_\odot$.
In contrast, LIRGs
and LLIRGs show more peaked distributions with
median values of $M_\star\approx 9\times 10^9M_\odot$ and
$M_\star\approx 2\times10^9M_\odot$. Notice that the median value
for the LLIRGs detected in the mid-IR is
significantly larger than the corresponding value for the mid-IR undetected LLIRGs
and is more similar to that of LIRGs, $M_\star\approx 1.1\times10^{10}M_\odot$.
This shows that the 24$\mu m$ cut
at low luminosity introduces a bias toward the more massive objects.

The age distribution of our galaxies is plotted in Fig. 10. Notice that the definition we
give here to age is different from that usually adopted when galaxy SEDs are analyzed by
means of a unique SSP. In this latter case one derives the luminosity-weighted mean age of
all the stellar populations present in the galaxy. In our case we refer to the age of the oldest
population present in the galaxy. This is a free parameter in the fit constrained only by a
maximum value, which is the Hubble time at the galaxy redshift. The lower limit to the ages
of ULIRGs and LLIRGs are not significantly different and, perhaps more importantly, both
classes contain very young objects, with ages of only a few tens of Myr. This is particularly
relevant for the ULIRGs because it is a strong and direct indication that a significant amount
of dust must already be in place after such a short timescale.

The median age of the sub-samples increases at decreasing IR luminosity. For ULIRGs
it is $\approx$150 Myr, for LIRGs it is $\approx$200 Myr,
and for LLIRGs it reaches $\approx$900 Myr. The
latter value decreases to $\approx$600 Myr if we consider
the LLIRGs undetected at 24 $\mu m$. The
distribution of ULIRGs is significantly flatter than that of LIRGs and in all cases objects as
old as about 2 Gyr have been found. In contrast, the distribution of the LIRGs and LLIRGs
is skewed toward significantly older ages. Given the meaning of our age, this actually means
that, on average, less luminous galaxies look older than more luminous ones. At first glance
this evidence could appear to be against the popular downsizing scenario, according to which
less massive galaxies should be younger (e.g.,  Noeske et al. 2007; Clemens et al. 2010).
We notice, however, that the downsizing scenario is
generally based on luminosity-weighted ages that, as already said, are averaged over all the
galaxy stellar populations. Thus a likely possibility to reconcile both observational results is
that the star formation in low luminosity galaxies started first but, having a lower efficiency
or being less suffering impulsive energy feedback (likely from quasars), can be maintained for
a longer time. Thus low mass objects would on average form first but they would appear
nowadays younger because their star formation lasted for a longer time. This interpretation
would be also fully consistent with the evidence that the partition of heavy elements in
local lower luminosity galaxies is progressively
less $\alpha$-enhanced (e.g., Clemens et al. 2010),
another fact that can be explained by an increasing duration of the star formation process.

Yet another possibility to explain the above observation is that the duty cycle of the
more massive objects is shorter both because of a shorter duration of star formation and of a rapid
increase in the dust content that renders these objects undetectable in the UV/optical after a
short timescale. In this case we will also preferentially observe younger objects.
Despite this, relatively older massive galaxies could also be included in the current sample
if their average attenuation decreases near the completion of their star formation history. In
any case, our findings are consistent with the predictions of a downsizing scenario induced
by an anti-hierarchical mass assembly process in which lower mass galaxies on average do
form first, but their star formation last for a longer period due to their inability to get rid
of the gaseous component (see Granato et al. 2004; Lapi et al. 2011). In the more massive
objects, instead the quasar feedback is able to remove the gas fueling in a shorter timescale
(see Lapi et al. 2006).

\begin{figure}
\centering
\includegraphics[width=0.8\textwidth,angle=0]{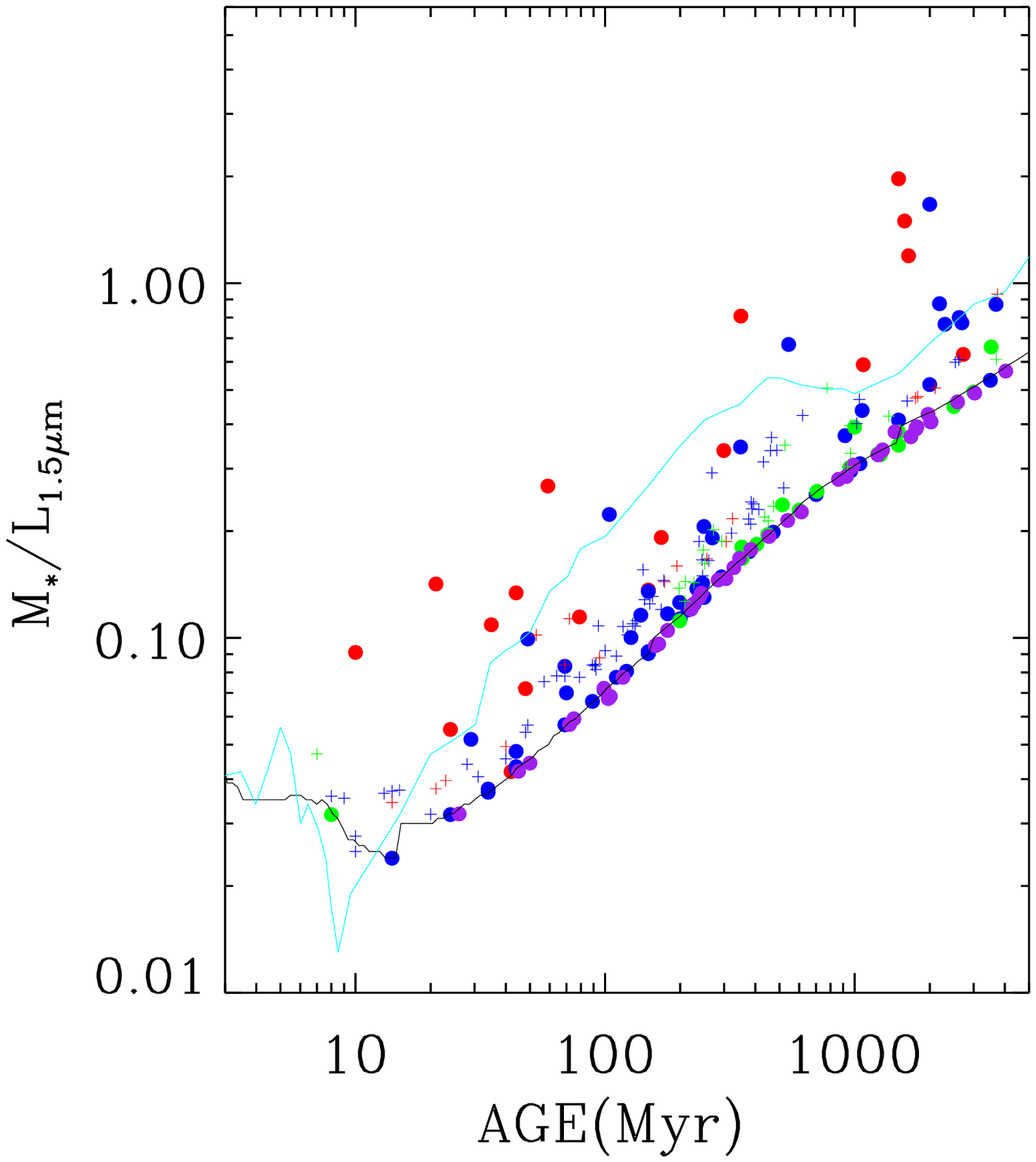}
\caption{Observed mass to light ratio in the rest-frame H-band vs. the galaxy age. Color-code is the same
as in Fig. 6. The luminosities are the observed ones, i.e., without dust correction.
The filled circles represent the results according to our SED-fitting technique (IR prior and
variable $R_V$ ), while the crosses refer to the standard technique
(no IR prior and $R_V = 4.05$).
The lines represents the locus of the unattenuated models, for a continuous SFR (black) and
a unique SSP (cyan).}
\end{figure}

Fig. 11 shows the specific star formation rate (sSFR=SFR/$M_\star$) as a function of the
stellar mass $M_\star$. We also plot the three loci with constant SFR = 10, $10^2$
and $10^3M_\odot yr^{-1}$,
from left to right. In a similar figure Reddy et al. (2006) showed that their galaxies lie in
a narrow band, suggesting an eventual relation of the sSFR with the mass of the galaxy.
However, their MIPS 24 $\mu m$ undetected objects occupy a lower region, indicating that the
above hint may result from a bias induced by the adopted selection criteria. Our selection
allows us to extend the range both at higher and at lower SFR, and we do not find any
noticeable trend of the sSFR with the mass of the galaxy. Actually, the sSFR may change
by about two order of magnitude for masses up to $M_\star\approx 10^{11}M_\odot$.

\subsection{Mass to Light ratios}

In Fig. 12 we plot the intrinsic (i.e., the luminosity is corrected for attenuation) mass
to light ratio in the rest-frame H band, in solar masses per solar H-band luminosity, vs.
the absolute H-band magnitude. In the left panel we plot the results obtained with our
procedure while in the right panel we show for comparison the results obtained by adopting the
standard procedure. The mass to light ratios span a factor of $\geq$30 from 0.02 to 0.6, in both
cases. The agreement is due to the fact that, for a constant SFR, the intrinsic $M_\star/L$ ratio
is only a function of age and the age distribution is similar (though the age of individual
objects is not exactly same).

However, we notice that in our case one cannot simply rely on the notion that the
attenuation in the near-IR is low. Especially in the brightest galaxies the attenuation in the
near-IR may be large, and neglecting this effect would produce significantly higher $M_\star/L$
ratio. To better clarify this point we plot in Fig. 13 the mass to light ratio versus the galaxy
age. The data points correspond to the observed luminosity while the solid line represents
the locus of the unattenuated models. We see that the intrinsic $M_\star/L$ ratios range from 0.02
to about 0.6 while, obviously, the uncorrected $M_\star/L$ ratio reach much higher values. The
observed spread at a given age is only due to attenuation.

Other authors claimed to have detected high $M_\star/L$ ratios in high redshift galaxies, even
consistent, in the near-IR passbands, with those of local old metal-rich systems (see Shapley
et al. 2005, 2011). Since they have used the standard procedure, where at these wavelengths
the attenuation is negligible, we conclude that the difference with respect to our finding is
simply due to the differences in the underlying stellar populations. As we already said, the
$M_\star/L$ ratios in models with constant SFRs depend only on the age and thus these claims
imply that, in the adopted models, the $M_\star/L$ ratios of intermediate age populations are
similar to those of old systems. Thus for a constant SFR model it is impossible to reach,
after a few of Gyr, the values of $M_\star/L\geq1.5$ observed in local passively evolving spheroids.

\section{Summary}

We have constructed a sample of 298 spectroscopically-confirmed galaxies at redshift
$z\sim2$, selected in the $\mz$-band from the GOODS-MUSIC catalog, with a SED well-sampled
from the UV to the IR. This allows the analysis of the galaxy physical parameters by
means of the popular SED-fitting technique. In doing this, we have adopted simple and
standard assumptions like constant SFR, solar metallicity, and age-independent attenuation.

The $\mz$-selection includes objects with a wide range of SFR, between a few to $10^3M_\odot yr^{-1}$,
and with different strengths of the attenuation. Since about half of the sample (135
objects) has been detected at 24 $\mu m$ with Spitzer and since it has been claimed that the rest
frame 8 $\mu m$ luminosity is a quite good proxy of the SFR, we have checked, for the first time
in an intermediate redshift sample, the accuracy of the standard SED fitting technique.

We find that the standard SED fitting technique is not accurate enough to provide reliable
estimate of the SFR and, correspondingly, of the attenuation, of the galaxy age and stellar
masses.

There is a large scatter in the predicted vs. expected IR luminosity (that can reach
even a factor of $\approx$ 50). The discrepancy increases at
increasing star formation and, for the
ULIRGs, the standard technique underestimates the SFR by even an order of magnitude.
Surprisingly, the discrepancy is not limited to the most obscured galaxies, as expected, but
it is also present in the low luminosity sample and, in some cases, the SED-inferred SFR
is even larger than that probed from the observed 24 $\mu m$ luminosity. The total mass is on
average underestimated by an average factor of about 50\%, but individual values may be
offset even by a factor of 4.

Combining the observed IR luminosity with the observed UV spectral slope we have also
constructed the rest frame $IRX-\beta_{UV}$ diagram for this redshift $z\sim2$ sample.
This diagram
is similar to other ones existing in literature and shows a large spread around the average
Meurer relation. This spread can be well reproduced by changing the ratio between the
neutral and the selective attenuation ($R_V$) in the Calzetti law. We interpret this fact as an
actual difference between an extinction and an attenuation laws, the latter being the result
of the combination of many effects, such as an age-dependent extinction, complex geometries
of dust and stars, and possible differences between grain compositions and sizes.

We have thus reanalyzed the 24 $\mu m$-detected subsample exploiting the constraint imposed
by the total IR luminosity and the possibility of varying the attenuation law.
Moreover, instead of adopting usual conversions from the IR to the SFR (see Kennicutt
1998; Panuzzo et al. 2003), that hold only in particular conditions (e.g., fixed duration of
the star formation episode, given geometry, etc.) we have used the observed IR luminosity
as a constraint in the SED-fitting technique.

As a result we have outlined a new method that should be used when the knowledge of
the IR luminosity can be added to the global SED. The most noticeable difference between
the methodology used in the present with respect to previous works rests on

\begin{itemize}
\item {\sl the use of the IR constraint, here derived from the observed 24 $\mu m$ luminosity;}

\item {\sl the use of a generalized Calzetti law with a variable $R_V$ .}
\end{itemize}

With this new method we have determined in an unprecedent robust way the physical
parameters that characterize our galaxies, namely SFR, attenuation and age.

\begin{itemize}
\item {\emph{Attenuation}.There is a general trend of increasing attenuation with SFR,
or, equivalently,
with the unobscured absolute UV magnitude, but the correlation is not tight.
The ratio $R_V$ shows a large scatter at all values of the SFR indicating a variety of
concomitant effects that may bear on the combined attenuation law. This was already
suggested by our $IRX -\beta_{UV}$ diagram but the run of the scatter with the SFR is more
indicative that these processes appear at any scale.

The relation between the attenuation and the SFR become tighter toward larger wavelengths.
This is consistent with the fact that in general at shorter wavelengths images
show more knotty structures that overall combine to give a different attenuation curve.
High values of $R_V$ were generally required to fit the SEDs of ULIRGs. In this case the
attenuation curve is very flat and the attenuation remains relatively significant even in
the near-IR. Indeed, from the analysis of local ULIRGs, it was expected that a fraction
of the star formation could be significantly attenuated even at these long wavelengths.
For the low luminosity galaxies the knowledge of the IR prior is not as important as
for those with high luminosity. This threshold corresponds to a SFR of about $20M_\odot
yr^{-1}$. However, if we use a pure Calzetti low for these galaxies, the standard procedure
slightly overestimates the SFR. A more suitable attenuation law is that with $R_V\approx3.0$.
There is a fair relation of the UV attenuation and the $\mz - 4.5 \mu m$ color. This could
eventually be used to obtain the absolute UV magnitude in absence of a IR luminosity
prior, but its effective usefulness has not been studied in detail, and is deferred to a
future work.}

\item {\emph{Star formation rates, ages and masses.}Our $\mz$-selected sample
includes objects with a
wide range of SFR, between a few to $10^3M_\odot yr^{-1}$. The average SFR is about 80$M_\odot yr^{-1}$.
In this respect the sample complements other ones selected by means of the
drop-out technique.

The mass distribution ranges from $10^9$ to $4\times10^{11}M_\odot$. If we take into account the
sample of low luminosity 24 $\mu m$ undetected sources there is a strong evidence that
the average mass increases at increasing star formation. ULIRGs show a flat mass
distribution with a minimum mass of $1.2\times10^{10}M_\odot$.

The age distribution of our sample ranges from a few tens of Myr to more than 1 Gyr.
For the first time we have obtained accurate ages of severely obscured, intermediate
redshift objects with very high SFRs. Individual age measurements of highly attenuated
objects indicate that dust must form within a few tens of Myr and be copious
already by the time the most massive AGB stars are evolved, i.e., at times $\leq$100 Myr.
Low luminosity, star-forming galaxies detected at 24 $\mu m$ show on average a significantly
more prolonged star formation with respect to more luminous star-forming objects,
though their mass is not significantly different. However, the 24 $\mu m$ detection threshold
is too heavy for this subsample and must be released in order to produce a statistically
significant number of objects. By releasing this constraint and using the attenuation
law with $R_V\approx3.0$ in the standard procedure, we could increase the subsample to 41
galaxies. As explained above, for low luminosity objects this method does not affect
the accuracy of the results.

With this increased sample we confirm that low luminosity galaxies harbor, on average,
significantly older stellar populations and are also less massive than the brighter ones.
Those selected at 24 $\mu m$ constitute actually the brightest tail of the low-luminosity
subsample. Thus we confirm that the observed downsizing effect (lower mass galaxies
appearing younger), is consistent with a picture where less massive galaxies actually
form first, but their star formation lasts longer, consistent with an anti-hierarchical
galaxy formation scenario.}

\item {\emph{Specific star formation rate.}
We do not find any trend of the sSFR with the mass mass
of the galaxy as claimed by other authors. We discuss how the previously observed
trend can be spurious and results from a bias induced by the selection criteria of the
analyzed samples.}

\item{\emph{Mass to light ratios.}
We find that care must be taken when dealing with $M_\star/L$ ratios
because, while it is customarily assumed that attenuation is scarce at near-IR
wavelengths, one remarkable result of our investigation is that this is not the case for
luminous objects. After imposing the IR constraint we find that, in order to fit the
whole SED, we need a flatter attenuation curve. This implies that the attenuation is not
negligible at near-IR wavelengths. Correspondingly, the near-IR $M_\star/L$ ratios obtained
after a proper attenuation correction never reach those of nearby galaxies, as claimed
by other authors, and remain lower by about a factor of three. In this respect the objects
with the higher $M_\star/L$ ratio are the low luminosity sources, because they are also the
oldest galaxies.}
\end{itemize}

\acknowledgments
This work has been supported by the Chinese National Science Foundation  (NSFC-11203023) and Chinese Universities Scientific Fund  (WK2030220011,WK2030220004,WJ2030220007). L.F. thanks the partly financial support from the
China Postdoctoral Science Foundation (Grant No.:2012M511411, 2013T60615). A.L. thanks partly support by ASI and MIUR.

\appendix

\section{Simple Stellar Population Models}

Simple stellar population (SSP) models adopted here are computed following Bressan et
al. (1998) but with the new empirical stellar spectral library MILES (S\'anchez-Bl\'azquez et al.
2006), which covers well the parameter space of metallicity, effective temperature and gravity
and thus ensures good optical broad-band starting colours. To cope with its spectral range
limitations we have extended the stellar spectra of MILES by means of matched NEXTGEN
(Allard et al. 2000) models both in the far-UV and in the near-IR/mid-IR spectral region.
For temperatures above 10000 K we do not have NEXTGEN models and we have used the
library by Munari et al. (2005) with the extension of the Lejeune models below $2500\AA$ and
above $10000 \AA$. In this way we have obtained three core libraries with average $[M/H] =-0.5$,
0.0 and $+0.3$. The stellar spectra extend from $10\AA$ to $160 \mu m$ and the spectral resolution
is of about $2\AA$ FWHM, from $2500\AA$ to $10000\AA$. To take into account the effects of mass
loss in hot supergiant stars we have considered two further extensions. For O stars we have
considered the spectral models by Schaerer \& de Koter (1997) while for Wolf Rayet stars we
have included the spectral models by Schmutz, Leitherer \& Gruenwald (1992). The latter
library provides only the continuum distribution of Wolf Rayet stars while there are more
recent stellar libraries that provide also spectral features (e.g., Smith et al. 2002). However,
the Schmutz et al. library is suitable for our purposes because we are considering only broadband
magnitudes here and, more importantly, because its parametrization as a function of
core temperature $T^\star$ and transformed radius $R_t$ allows a fair interpolation between stellar
evolution quantities ($L$ and $T_{eff}$) and spectral models of thick winds.

For the young populations we have also considered the nebular spectrum, which is calculated
by means of \texttt{CLOUDY} (Ferland 1996) assuming case B recombination. To compute
the nebular emission at different ages we have considered the corresponding spectra of the
SSPs using the following parameters: mass of the ionizing cluster $10^5M_\odot$, electron number
density $n = 100 cm^{-3}$, inner radius 15 pc, and the metallicity rescaled to that of the SSP.
The nebular emission has been then suitably rescaled to the original mass of the SSP. The
main effect we noticed on broad-band fluxes is the contribution of nebular continuum in very
young stellar populations. Changing the nebular parameters would not affect our conclusions
significantly.

For intermediate-age stellar populations we have also considered the effects of dusty
envelopes around asymptotic giant branch stars (AGBs), as described in Bressan et al.
(1998), but after revisiting the mass loss rate formulation and the expected dust composition,
as described in Marigo et al. (2008). We have finally checked that the isochrones reproduce
well the observed integrated (V-K) colours of LMC star clusters (Persson et al. 1983; Kyeong
et al. 2003; Pessev et al. 2006), especially at intermediate ages.

\label{lastpage}
\end{document}